\documentclass[journal=jpcc,manuscript=article]{achemso}

\usepackage[version=3]{mhchem} 
\usepackage[final]{pdfpages}

\author{Esteban~Rojas-Gatjens}
\affiliation{School of Chemistry and Biochemistry, Georgia Institute of Technology, Atlanta, GA 30332, United~States}

\author{Kaila~Yallum}
\affiliation{Department of Chemistry, Biochemistry, and Pharmaceutical Sciences, University of Bern, Freiestrasse 3, CH-3012 Bern, Switzerland}

\author{Yangwei~Shi}
\affiliation{Department of Chemistry, University of Washington, Seattle, WA 98195, United~States}
\alsoaffiliation{Molecular Engineering \& Sciences Institute, University of Washington, Seattle, WA 98195, United~States}

\author{Yulong~Zheng}
\affiliation{School of Chemistry and Biochemistry, Georgia Institute of Technology, Atlanta, GA 30332, United~States}

\author{Tyler~Bills}
\affiliation{School of Chemistry and Biochemistry, Georgia Institute of Technology, Atlanta, GA 30332, United~States}

\author{Carlo~Andrea~Riccardo~Perini}
\affiliation{School of Materials Science and Engineering, Georgia Institute of Technology, Atlanta, GA 30332, United~States}

\author{Juan-Pablo~Correa-Baena}
\affiliation{School of Materials Science and Engineering, Georgia Institute of Technology, Atlanta, GA 30332, United~States}

\author{David~S.~Ginger}
\affiliation{Department of Chemistry, University of Washington, Seattle, WA 98195, United~States}

\author{Natalie~Banerji}
\affiliation{Department of Chemistry, Biochemistry, and Pharmaceutical Sciences, University of Bern, Freiestrasse 3, CH-3012 Bern, Switzerland}

\author{Carlos~Silva-Acu\~na}
\email{carlos.silva@gatech.edu}
\affiliation{School of Chemistry and Biochemistry, Georgia Institute of Technology, Atlanta, GA 30332, United~States}
\alsoaffiliation{School of Physics, Georgia Institute of Technology, Atlanta, GA 30332, United~States}
\alsoaffiliation{School of Materials Science and Engineering, Georgia Institute of Technology, Atlanta, GA 30332, United~States}

\title{Resolving nonlinear recombination dynamics in semiconductors via ultrafast excitation correlation spectroscopy: Photoluminescence versus photocurrent detection}

\begin{document}
\begin{tocentry}
    \centering
    \includegraphics[width=0.65\linewidth]{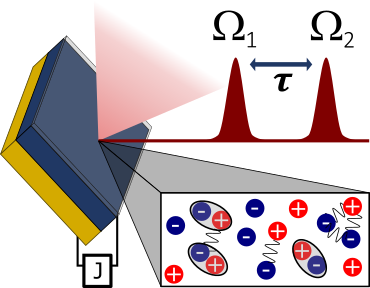}
\end{tocentry}

\newpage
\begin{abstract}
We explore the application of excitation correlation spectroscopy to detect nonlinear photophysical dynamics in two distinct semiconductor classes through time-integrated photoluminescence and photocurrent measurements. In this experiment, two variably delayed femtosecond pulses excite the semiconductor, and the time-integrated photoluminescence or photocurrent component arising from the nonlinear dynamics of the populations induced by each pulse is measured as a function of inter-pulse delay by phase-sensitive detection with a lock-in amplifier. 
We focus on two limiting materials systems with contrasting optical properties: a prototypical lead-halide perovskite (LHP) solar cell, in which primary photoexcitations are charge photocarriers, and a single-component organic-semiconductor diode, which features Frenkel excitons as primary photoexcitations. The photoexcitation dynamics perceived by the two detection schemes in these contrasting systems are distinct. Nonlinear-dynamic contributions in the photoluminescence detection scheme arise from contributions to radiative recombination in both materials systems, while photocurrent arises directly in the LHP but indirectly following exciton dissociation in the organic system. Consequently, the basic photophysics of the two systems are reflected differently when comparing measurements with the two detection schemes. Our results indicate that photoluminescence detection in the LHP system provides valuable information about trap-assisted and Auger recombination processes, but that these processes are convoluted in a non-trivial way in the photocurrent response and are therefore difficult to differentiate. In contrast, the organic-semiconductor system exhibits more directly correlated responses in the nonlinear photoluminescence and photocurrent measurements, as charge carriers are secondary excitations only generated through exciton dissociation processes. We propose that bimolecular annihilation pathways mainly contribute to the generation of charge carriers in single-component organic semiconductor devices. 
Overall, our work highlights the utility of excitation correlation spectroscopy in modern semiconductor materials research, particularly in the analysis of nonlinear photophysical processes, which are deterministic for their electronic and optical properties.
\end{abstract}

\section{\label{Intro}Introduction}

Probing photoexcitation dynamics is a cornerstone of materials characterization in optoelectronics. 
Photoexcitations may undergo radiative recombination, defect trapping/detrapping, and high-order processes such as bimolecular and Auger recombination, among others. 
These recombination dynamics often give rise to a nonlinear response with respect to photoexcitation density, and in turn, they can be determined by resolving time-dependent populations. 
Researchers commonly probe the nonlinear response using intensity-dependent (i) steady-state photoluminescence and photocurrent experiments, in which a deviation of the signal $S(I)$ from a linear response, $S(I) \propto I^{\alpha}$ is observed~\cite{VanOrman2021}, (ii) time-resolved photoluminescence~\cite{Herz2004, Riley2022} or (iii) transient absorption spectroscopies~\cite{Silva2001, Clement2007, Firdaus2020}. 
However, delimiting the distinct nonlinear regimes can be ambiguous between these techniques, and may become complex as the system's components increase. 
Originally described by von der Linde and Rosen~\cite{VONDERLINDE1981, Rosen}, excitation correlation (EC) spectroscopy provides the means to characterize the nonlinear response with great sensitivity as it is based on double amplitude modulation and phase-sensitive detection. 
In addition, it maps the time evolution of nonlinear contributions. 
In EC spectroscopy, we amplitude-modulate two replica ultrafast pulses at frequencies $\Omega_1$ and $\Omega_2$, such that demodulating at a reference frequency $|\Omega_1+\Omega_2|$ using lock-in detection isolates the nonlinear component. 
EC spectroscopy has been widely used to characterize the carrier lifetimes of several inorganic semiconductors~\cite{Johnson1988, Chilla1993, pau1998, Hirori2006, Miyauchi2009}.
Despite its utility, neither the organic nor the lead halide perovskite (LHP) semiconductor community uses it as a routine technique. 
Only recently, Srimath~Kandada~\textit{et al.} employed it to describe the defect density and energetic depth in \ce{CH3NH3PbBr3} thin films and \ce{CsPbBr3} nanocrystals~\cite{Kandada2016}.
The Moran group presented a variation of the EC spectroscopy, utilizing a tunable narrow excitation wavelength to characterize layered perovskite quantum-well structures~\cite{Zhou2019, Ouyang2020, Zhou2021}.  
We have previously implemented ECPL to describe defect states in mix-halide mix-cation metal halide perovskites~\cite{Perini2022, Yangwei2022}.

In this article, we implement EC spectroscopy with both photoluminescence (PL) and photocurrent (PC) detection to characterize the nonlinear response of two photodiodes, a LHP in a solar cell, and an organic semiconductor single-component vertical diode. 
We describe in detail the interpretation of EC signatures using simplified kinetic recombination models that exemplify the class of nonlinear dynamics in these materials systems.  
First, we discuss the typical photophysical processes that result in ECS signals for the case of LHP. 
We show that trap-assisted and Auger recombination dominate the nonlinear response of LHP devices in PL detection (ECPL) at low and high fluence, respectively. In PC detection (ECPC) the nonlinear components are due to bimolecular and Auger recombination, however, these contributions cannot be easily distinguished with this detection scheme. 
We also describe the photophysical scenario of the organic semiconductor diode leading to ECS signal. In this case, where the primary excitation is a Frenkel exciton, and charge carriers are not directly injected, ECPL and ECPC provide complementary information about the population evolution of excitons and charges. 

\section{\label{Res}Results}

\begin{figure}
    \centering
    \includegraphics[width=0.5\linewidth]{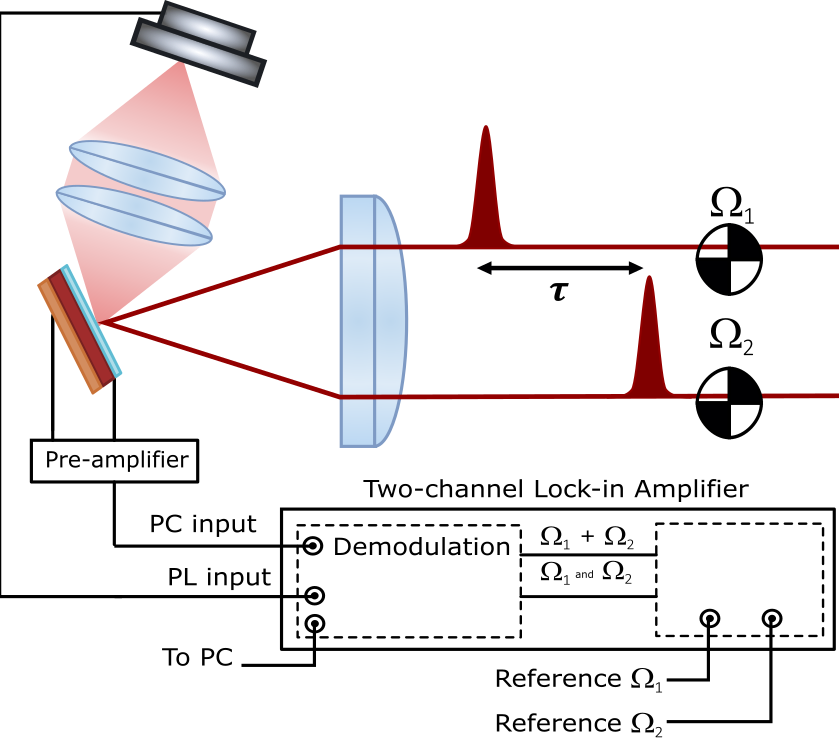}
    \caption{Schematic representation of the excitation correlation measurement. The photoluminescence signal is measured using a photodiode and the photocurrent is processed by a current amplifier. Both signals are sent to the lock-in amplifier, which demodulates the input signal at the fundamental of the two amplitude modulation reference waveforms with frequencies $\Omega_1$ and $\Omega_2$, and at the sideband $\Omega_1 + \Omega_2$.}
    \label{fig:my_schem}
\end{figure}

\subsection{Nonlinear dynamics in lead-halide perovskites} \label{SectionPerov}

We prepared inverted devices with a mixed-cation mixed-halide perovskite of composition \ce{FA_{0.83}Cs_{0.17}Pb(I_{0.85}Br_{0.15})_3} referred to in the text as Cs17Br15, where \ce{FA+} refers to the formamidinium cation. Fig.~S1(a), in Supplementary Material, shows the linear absorption and PL spectra of a Cs17Br15 film on ITO/MeO-2PACz.  Our supplementary material and previous work~\cite{Yangwei2022} provide further details on the device structure and characterization. Briefly, these films and fabrication procedures yield performance of around 15.90\% power conversion efficiency for the best devices. The external quantum efficiency (EQE) measurement is shown in Fig.~S2 in Supplementary Material. 
We emphasize that we perform both ECPL and ECPC on completed device stacks, with typical PL quantum yields in the range of 0.8\%. 
To perform ECPL and ECPC, we excite the sample with a 220\,fs pulse with an energy of 2.638\,eV and variable fluence between 1 and 40\,$\mu$J\,cm$^{-2}$. 
A schematic representation of the EC experiment is represented in Fig.~\ref{fig:my_schem}, and more details about our implementation are described in the appendix. 

\begin{figure}
\includegraphics[width=0.5\columnwidth]{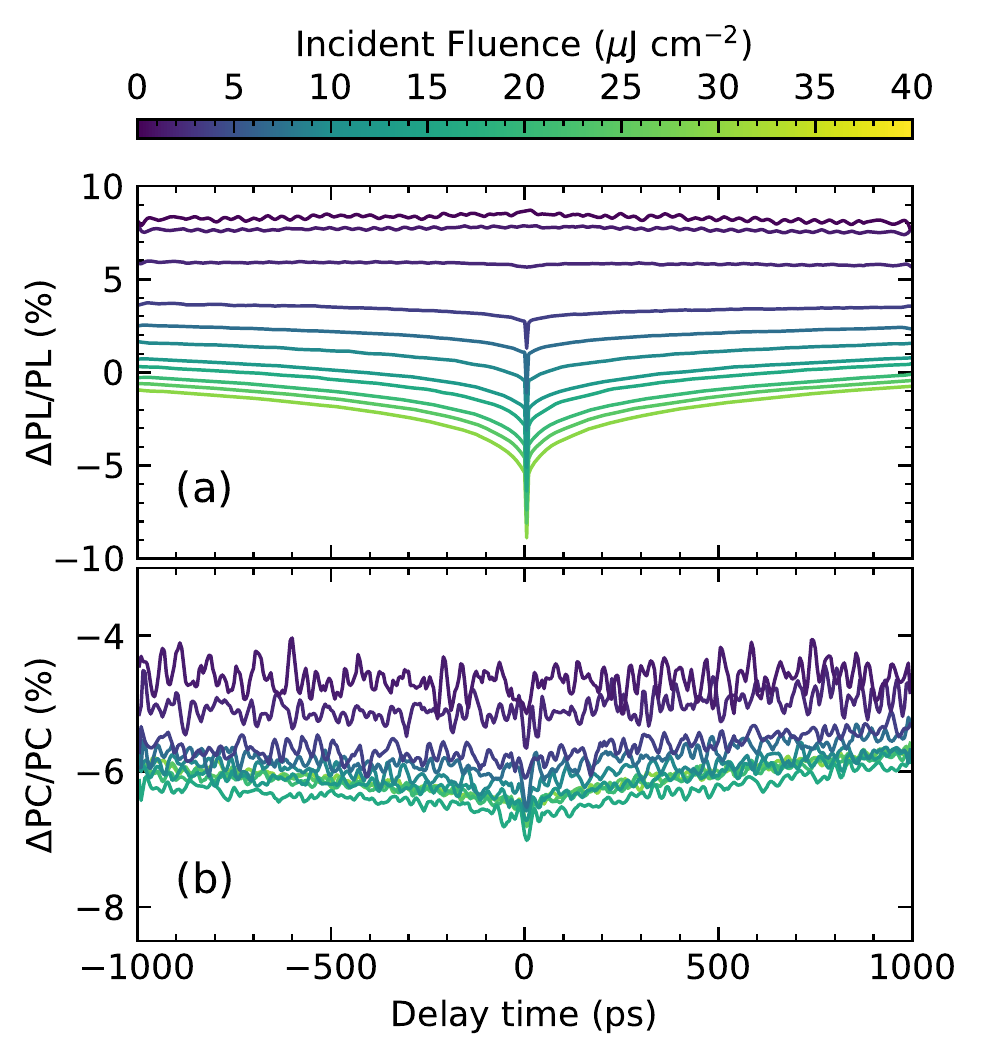}
\caption{Excitation correlation spectroscopy measurement of a prototypical Cs17Br15 device. (a) PL detected and (b) PC detected nonlinear response. In both cases, the pump wavelength was 470 nm, and the fluence range is indicated in the false color axis.}
\label{fig:PEROV}
\end{figure}

Because metal-halide hybrid perovskites are direct bandgap semiconductors, their recombination kinetics involve photocarriers undergoing second-order (bimolecular) radiative recombination of electrons and holes, pseudo-first order radiative recombination of photogenerated minority carriers with the majority carriers, first-order deep-trap assisted non-radiative recombination, and third-order Auger recombination~\cite{Stranks2014, deQuilettes2019, Kiligaridis2021}. 
These terms are well described by equations~\ref{n}, \ref{p} and \ref{nt}, where $B$ is the bimolecular rate constant, $\gamma_t$ the carrier trapping rate constant, $\gamma_r$ the trap recombination rate constant, and $\gamma_{Auger}$ the Auger recombination rate constant. 
Additionally, $N_t$ and $n_t$ correspond to empty and occupied trap sites. 
The generation of electrons and holes is assumed to be direct, then their generation populations are considered as initial conditions when solving the differential equations.
Note that we do not take into account non-geminate association and dissociation of excitons explicitly since it does not add a distinct recombination order, and additionally, excitons are not generated, nor stable at room temperature~\cite{Ziffer2016}. 
%
\begin{equation}
    \frac{dn}{dt} = -Bnp-\gamma_t(N_t-n_t)\,n-\gamma_{Auger_1}pn^2 - \gamma_{Auger_2}p^2n,
    \label{n}
\end{equation}
\begin{equation}
    \frac{dp}{dt} = -Bnp - \gamma_r n_t p - \gamma_{Auger_1}pn^2 - \gamma_{Auger_2}p^2n,
    \label{p}
\end{equation}
\begin{equation}
    \frac{dn_t}{dt} = \frac{dn}{dt}-\frac{dp}{dt}
    \label{nt}.
\end{equation}
A distinct model assuming shallow donors has been used to describe \ce{CH3NH3PbBr3}~\cite{Kandada2016}, where shallow traps dope the semiconductor. Based on this model, we expect to observe positive subnanosecond dynamics due to fast trapping in shallow traps, accompanied by an increase in the ECPL response as the excitation fluence increases. 
However, this model does not apply to the Cs17Br15 devices in our study, as discussed below. 
The ECPL and ECPC measurements are shown in Fig.~\ref{fig:PEROV}(a) and Fig.~\ref{fig:PEROV}(b), respectively. 
The time traces show the percentage of the nonlinear signal recovery as we scan the delay between the two pulses, the symmetry between the negative and positive delays indicates that the pulses had a comparable intensity. 
For the ECPL, At low fluence, we observe a nonlinear response with slow dynamics. 
The magnitude of the nonlinear response decreases as the fluence increases until it changes sign at the highest fluences. 
We rule out the shallow donor model as these experimental signatures do not match the model's prediction that Kandada\,\textit{et al.} described in Ref~\citenum{Kandada2016}.
The ECPL signal in Fig~\ref{fig:PEROV}(a), shows two distinct regimes: at low fluence, a slow positive trace, and at high fluence, a fast negative nonlinearity. 
In contrast, the ECPC response, shown in Fig.~\ref{fig:PEROV}(b), shows only negative contributions, with no change in the sign of the signal as the fluence increases. 
To rationalize the information provided by each technique, we interpret the EC measurements of LHPs in terms of the recombination model. 
Equations~\ref{n}, \ref{p} and \ref{nt} do not have an analytical solution. 
However, by making a series of assumptions described below we can understand the contributions of the specific processes to the nonlinear photoluminescence and photocurrent. 

\subsubsection{Trap-Assisted Recombination}

We follow the assumptions made by previous works on trap-assisted recombination~\cite{Borgwardt2015, Johnson1988}. Specifically, we assume that we are working at low excitation density, such that Auger recombination does not dominate and can be neglected. Additionally, in materials with low PL quantum yield, as is the case for LHPs, nonradiative trap-assisted pathways typically dominate the carrier recombination such that $Bnp \ll \gamma_t N_t n$.
Consequently, there is an approximate solution for the electron and hole densities, which reads as:
\begin{align}
    n(t) = n(0)\exp(-\gamma_t N_t t)\label{np0},\\ 
    p(t) = p(0)\exp(-\gamma_r n_t t).
    \label{np1}
\end{align}

We will discuss first the case of ECPL. The detected PL is defined in equation~\ref{TotalPL}.
The temporal function describing carriers generated by each pulse is assumed to be a delta function.
We then split the integral describing the total PL intensity in two: one term considering the carriers photoexcited before the second pulse and another one with carriers photoexcited after the second pulse. Here, $n_1(t)$ and $p_1(t)$ correspond to the evolution of the carriers according to equations~\ref{np0}-\ref{np1} after the first pulse arrives, the initial conditions are simply the carriers generated by the first pulse, $n_1(0) = p_1(0) = n_0$. $n_2(t)$ and $p_2(t)$ correspond to the evolution of the carriers after the second pulse. These carrier densities are described with the same expression but with distinct initial conditions, $n_2(0)= n_0+n_1(\tau)$ and $p_2(0)= n_0+p_1(\tau)$, as we need to consider residual carriers generated by the first pulse.
\begin{equation}
    I_{Total\,PL} \propto \int_0^\tau n_1(t)p_1(t)dt+\int_\tau^\infty n_2(t-\tau)p_2(t-\tau)dt.
    \label{TotalPL}
\end{equation}
After integrating equation~\ref{TotalPL} and subtracting the individual pulse contributions ($2n_0^2$), we obtain the nonlinear component of the photoluminescence intensity ($I_{NPL}$) given in equation~\ref{NonlinearPL}. 
Note that the nonlinear term has a positive value, as expected since trap-filling results in a reduction of non-radiative decay pathways.
We note that under these assumptions the $I_{NPL}$ follows the same decay as conventional time-resolved experiments. 
Experimentally, in the ECPL measurements at low fluence shown in  Fig.~\ref{fig:PEROV}(a), we observe a slow decay rate, which is not entirely captured in the time window of the experiment. This is consistent with equation~\ref{NonlinearPL}, as the typical values for carriers' lifetimes in LHPs are between the nanosecond and microsecond range. The supplementary material shows the time-resolved photoluminescence experiments in Fig.~S3.
\begin{equation}
    I_{NPL} \propto \left(\exp(-\gamma_t N_t \tau)+\exp(-\gamma_r n_t \tau)\right).
    \label{NonlinearPL}
\end{equation}

We performed a similar analysis for the case of photocurrent detection. The signal measured is defined by equation~\ref{TotalPC}. 
We ignore the spatial distribution of the carriers and the extraction of carriers for the sake of simplicity. 
These assumptions affect the magnitude of the nonlinear signal.
We interpret the nonlinear photocurrent arising from carriers' interactions.
The time-resolved PL (see Fig~S3) indicates that the recombination kinetics in open-circuit and short-circuit conditions are very similar. Therefore, we justify using the same photophysical scenarios to interpret ECPL and ECPC. 
It is worth remembering that in ECPC the time resolution arises from the delay between the pulse delays instead of from the carrier device extraction.
Consequently, we only need the device charge extraction to be faster than the modulation frequency, which is the case by several orders of magnitude. 
\begin{equation}
    I_{Total\,PC} \propto \int_0^\tau \left( n_1(t)+p_1(t) \right)dt+\int_\tau^\infty \left( n_2(t-\tau)+p_2(t-\tau) \right )dt.
    \label{TotalPC}
\end{equation}
Under the assumption that trap-assisted recombination dominates at low fluences, the integrands correspond to the monoexponential decay equations~\label{np}. This is a linear function with the excitation density, therefore the nonlinear photocurrent is zero.
Trap-assisted recombination does not result in a nonlinear PC response, making ECPC insensitive to traps. 
However, in the experimental ECPC measurements in Fig.~\ref{fig:PEROV}(b), the nonlinear component at low fluence is not zero, and it has a negative value. 
Therefore, as discussed below, higher-order processes such as bimolecular recombination and Auger recombination must be responsible for the observed nonlinear photocurrent.

\subsubsection{Bimolecular Recombination}

We now consider the case where bimolecular recombination is the dominant recombination pathway. In the Supplementary Material, we show that if $Bnp \gg \gamma_tN_tn$, then the ECPL response is zero. This approach of ignoring completely the monomolecular recombination does not give an expression for ECPC as the integrals diverge. To attain an approximate analytical expression for the ECPC response, we assume that both bimolecular recombination, $B$, and monomolecular trapping, $\gamma = \gamma_t(N_t-n_t)$, are present. Also, we assume that holes and electrons have similar trapping rates such that $n \approx p$. This scenario is described by equation~\ref{bimolecular}, and the corresponding solution for the population evolution is shown in equation~\ref{nbi}.
\begin{equation}
    \frac{dn}{dt} = -Bn^2-\gamma\,n.
    \label{bimolecular}
\end{equation}
\begin{equation}
    n(t) = \frac{n_0\gamma/B}{(n_0+\gamma/B)\exp(\gamma t)-n_0}.
    \label{nbi}
\end{equation}

After integrating equation~\ref{TotalPC} and subtracting the individual pulses contribution ($4n_0$), we obtain an expression (equation~\ref{PCBI}) that describes the nonlinear photocurrent, $I_{NPC}$, where $\alpha = n_0B/\gamma$. 
Note that in the limiting cases where there is no bimolecular recombination ($B=0$), the nonlinear contribution to PC is zero, and at a long delay ($\tau$), the expression also goes to zero as the pulses do not overlap in time. Consequently, we can conclude that the nonlinear photocurrent arises from bimolecular recombination and not from carrier trapping, but the time evolution follows the carrier trapping dynamics. 
\begin{equation}
    I_{NPC} \propto \ln\left(1-\frac{\alpha^2\exp(-\gamma \tau)}{(1+\alpha)^2}\right) \approx -\frac{\alpha^2\exp(-\gamma \tau)}{(1+\alpha)^2}.
    \label{PCBI}
\end{equation}
According to equation~\ref{PCBI}, the nonlinear PC must have a negative sign, congruent with the experimental results shown in Fig.~\ref{fig:PEROV}(b). 
The ECPC measurements, similarly to ECPL, show slow time dynamics, which is expected as they both follow the time evolution of the carrier population. Additionally, note that in this scenario, the ratio between the bimolecular recombination and the carrier trapping rates dictates the magnitude of the nonlinear PC component.

We have neglected the spatial aspect of the carrier dynamics, which is relevant as we excite the sample in a small area of the sample.
Carrier dynamics simulations considering carrier diffusion, carrier trapping, and bimolecular recombination, have been carried out by Zhou \textit{et al.}~\cite{Zhou2021} for perovskite quantum wells.
They observe negative decaying nonlinear photocurrent at longer times, congruent with the slow traces shown in Fig.~\ref{fig:PEROV}(b).
So far, we have rationalized the slow dynamics and the sign of the ECPL and ECPC response at low fluences. 
Experimentally, as we increase the fluence, we observe a change in sign in the ECPL signal (Fig. 2(a)), while the ECPC response increases in magnitude but remains negative. As the fluence increases, Auger recombination becomes more significant and dominates the recombination kinetics. 
We will now rationalize the effect of Auger recombination in the nonlinear photoluminescence and photocurrent.

\subsubsection{Auger Recombination}

We next consider the scenario in which the carrier recombination is dominated by Auger scattering, a third-order process that occurs at high carrier density. Again, assuming that $p\approx n$ holds in the high-fluence regime, we describe the kinetics using the rate equation~\ref{Auger}. The solution of this equation is presented in equation~\ref{AugerPOP}. Here,  $\gamma$ corresponds to the monomolecular recombination rate constant, and $A$ corresponds to the Auger recombination rate constant.
\begin{equation}
\frac{dn}{dt} = -\gamma n - An^3.
\label{Auger}
\end{equation}
\begin{equation}
    n(t) = \sqrt{\frac{\gamma/A}{(1+\gamma/n^2_0A)e^{2\gamma t}-1}}.
    \label{AugerPOP}
\end{equation}
In this particular case, the expressions for the nonlinear photoluminescence and photocurrent are more complex to evaluate than in the previous scenarios. 
In the Supplementary Information, we show that both the photoluminescence and photocurrent exhibit negative nonlinear components due to Auger recombination. 
This negative contribution explains the fluence-dependent features that we observe for both ECPL and ECPC. 
In ECPL, when we transition from a trap-dominated recombination scenario with a positive nonlinear component to an Auger-recombination-dominated scenario with a negative nonlinear component, a change of sign is expected.
This transition is not expected in ECPC, as both recombination processes (bimolecular and Auger recombination) that result in nonlinear signals lead to negative nonlinear components, making it difficult to distinguish between the two scenarios. 
Additionally, we note that the ECPL experiments, Fig.~\ref{fig:PEROV}(a), show subnanosecond dynamics at the highest fluence (i.e. the sharp decay around zero time delay). 
We assign these fast features to a large population of carriers recombining through Auger pathways at early times. \\ 

\subsubsection{Summary of nonlinear dynamics in lead halide perovskites}
In summary, we have described the nonlinear responses caused by the distinct photophysical processes for both photoluminescence and photocurrent. We highlight the possibility of distinguishing between trap-assisted and Auger-dominated recombination regimes employing ECPL, as the contributions to the nonlinear response have opposite signs. The excitation density at which the ECPL signal changes sign indicates a change in the dominant process, which is related to the number of traps, meaning that ECPL is a good technique to characterize trap densities in LHPs. On the other hand, in ECPC, the nonlinear signal arising from bimolecular recombination and Auger recombination have the same sign (negative), monomolecular recombination by itself does not result in a nonlinear signal. Therefore, ECPC does not provide as rich information about trap density as ECPL since the signatures are convoluted and difficult to isolate. Recent work by Zhou \textit{et al.}~\cite{Zhou2019, Zhou2021} explores the implementation of a similar experimental setup to characterize carrier diffusion, this idea is not expanded in this work as large time scales (tens of nanoseconds), that exceed our implementation capabilities, are needed.

\subsection{Nonlinear dynamics in organic semiconductors} \label{2OSC}

In organic semiconductors, the primary photoexcitation is a neutral exciton. 
Charge carriers are generated after the dissociation of the exciton, which can occur through several mechanisms in the \emph{neat} semiconductor. 
One such mechanism is the formation of an intermediate charge transfer state prior to charge separation~\cite{Paquin2011}, although the precise mechanism for this process is not clear and is certainly not trivial. Another mechanism for exciton dissociation is to overcome the exciton binding energy by promoting the exciton to a higher energy excited state, $S_n^*$. 
This can be achieved by coherent two-step photo-excitation pathways using femtosecond-pulse excitation, described as an excitation from $S_0$ to $S_1$ and subsequently from $S_1$ to $S_n^*$. 
The process generates a high-energy state prone to relaxation to charged excitations (polarons) and triplet-excitons~\cite{Silva2001,Stevens2001,silva2002exciton}.
Alternatively, the $S_n^*$ state can be reached through energy transfer between excitons in a process known as exciton-exciton annihilation~\cite{Silva2001,Stevens2001,silva2002exciton, kohler2015electronic}, this is the mechanism proposed for this work, as discussed below.

The two detection methods used in the EC spectroscopy presented here are each sensitive to different excited-state species produced optically in organic semiconductors. 
While the charge carriers were both the emissive species and the PC-detected species in lead-halide perovskites, excitons and charges can be observed individually in the neat ITIC-4F devices studied here. 
Excitons correspond to the detected emissive species (Fig.~S1(b)in Supplemental Material), and charges, arising from subsequent exciton dissociation, result in the detected PC.
The ECPL experiment then provides insights into the processes leading to exciton recombination, while ECPC provides information on those resulting in charge-carrier generation. 
In this work, we assess the photophysical processes occurring in neat ITIC-4F.
We prepared a single component device with an architecture ITO/ZnO/ITIC-4F/\ce{MoO_3}/Ag, and measured both ECPL and ECPC. 
The absorption and PL emission spectra of ITIC-4F are shown in Fig.~S1(b) in Supplemental Material.  
Further details about the device preparation are presented in the Supplementary Material. 
The pump pulse used for these EC spectroscopy experiments has an energy of 1.823\,eV.

\subsubsection{Nonlinear photoluminescence} 

\begin{figure}
\includegraphics[width=0.5\columnwidth]{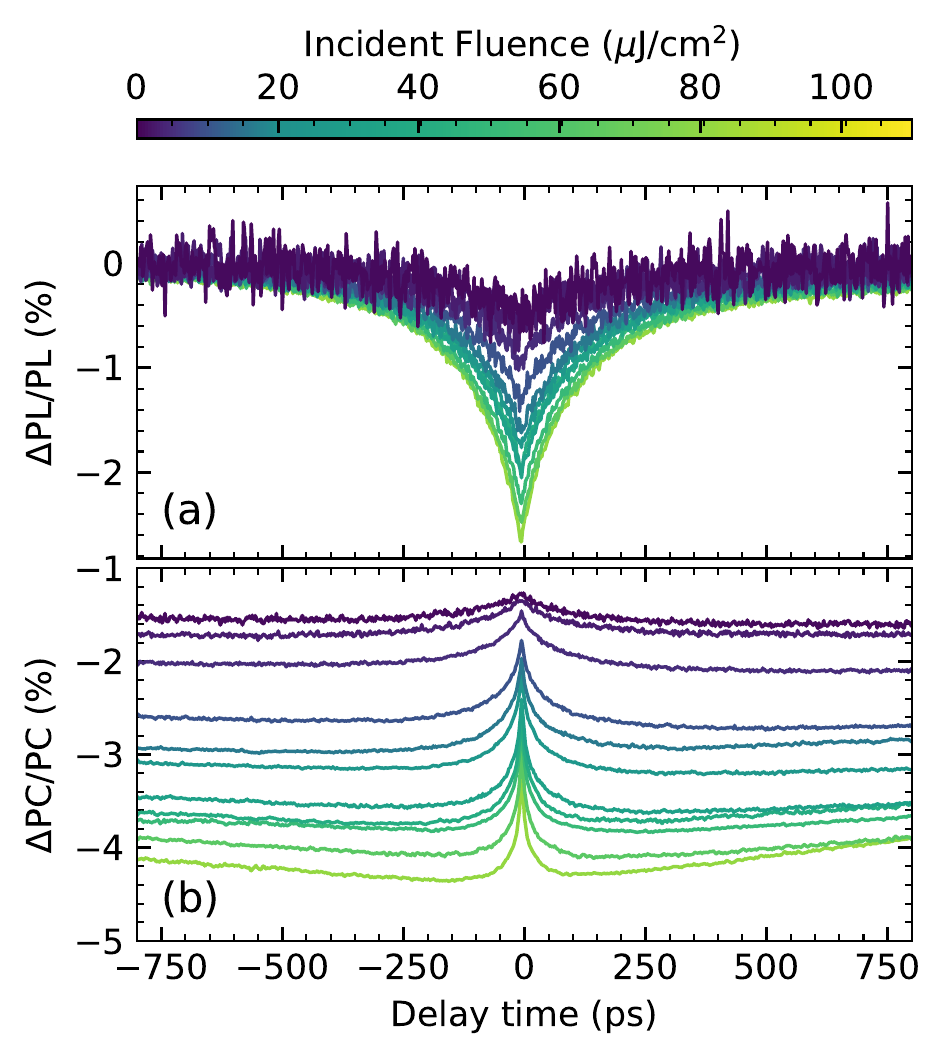}
\caption{Excitation correlation spectroscopy measurement of a ITIC-4F device. (a) Photoluminescence detected and (b) photocurrent detected nonlinear photocarrier dynamics. The pump wavelength was 680\,nm, and the fluence range is represented by the false color axis.}
\label{ECPLOrg}
\end{figure}

Consider the simple model in equation~\ref{Anh}, where the monomolecular rate $\gamma$ incorporates all monomolecular processes, including radiative and non-radiative relaxation pathways, and $\beta$ is the bimolecular exciton annihilation rate.
This model is mathematically equivalent to the equation~\ref{bimolecular} discussed for LHPs.
As shown above, equation~\ref{Anh} can be solved analytically and leads to the expression for the nonlinear photoluminescence shown in equation~\ref{bimol}. We define $\alpha = n_0\beta/\gamma$ equivalently.
\begin{equation}
    \frac{dn_{exc}}{dt} = -\gamma\,n_{exc}-\beta\,n_{exc}^2.
    \label{Anh}
\end{equation}
\begin{equation}
    I_{NPL} \approx \ln\left(1-\frac{\alpha^2\exp(-\gamma \tau)}{(1+\alpha)^2}\right).
    \label{bimol}
\end{equation}

Fig.~\ref{ECPLOrg}(a) shows the experimental ECPL response for a range of fluences between 1 and 100\,$\mu$J\,cm$^{-2}$. It can be observed that the magnitude of the nonlinear response increases with fluence as expected since exciton-exciton annihilation becomes more significant as the exciton density increases. 
By fitting the measured traces to equation~\ref{bimol}, we extracted a value of monomolecular recombination rate of $(6.3\pm 0.5) \times 10^9$\,s$^{-1}$. 
We extract the bimolecular rate using the experimental setup discussed by Riley~\textit{et al.} in Ref.~\citenum{Riley2022} and obtained a value of $(1.0\pm0.2)\times10^{-9}$\,cm$^3$\,s$^{-1}$, similar to those reported previously in the literature~\cite{Riley2022}.
A summary of the analysis and fitted data are shown in  Supplementary Material. 
In the experimental configuration of Ref.~\citenum{Riley2022}, the material is excited with a single pulse whose amplitude is modulated by a mechanical chopper. 
The bimolecular exciton-exciton rate was obtained after analyzing the fluence dependence of photoluminescence intensity.

In this section, by using a simple model including monomolecular and exciton-exciton recombination, we show that ECPL follows the excitons population time evolution through the nonlinear photoluminescence generated in organic semiconducting materials. 
More complex scenarios involving coexisting excitonic species have been analyzed previously~\cite{pau1998, Miyauchi2009}. 
In those cases, the additional exciton dynamics afflict the spectral integrated response, as the one measured in our work. Spectrally resolving the signal is necessary to distinguish nonlinear dynamics from distinct emissive species with similar emission energy.

\subsubsection{Nonlinear photocurrent} 

To interpret the photocurrent response in organic semiconductors, we will focus on a simple model for charge carrier population dynamics described by equation~\ref{last}. In this model, the photocarriers are generated through the function $G(t)$ which depends on the photophysical process that results in charges.
We acknowledge that the generation of photocarriers in neat organic semiconductors, and a complete description of their dynamics, is a complex problem and that multiple techniques are needed to provide a robust physical picture. 
However, in this work, we focus on the contributions that ECS can bring to the field, and thus we provide our hypothesis of the photophysical scenario in this materials class. 

\begin{figure}
\includegraphics[width=0.5\columnwidth]{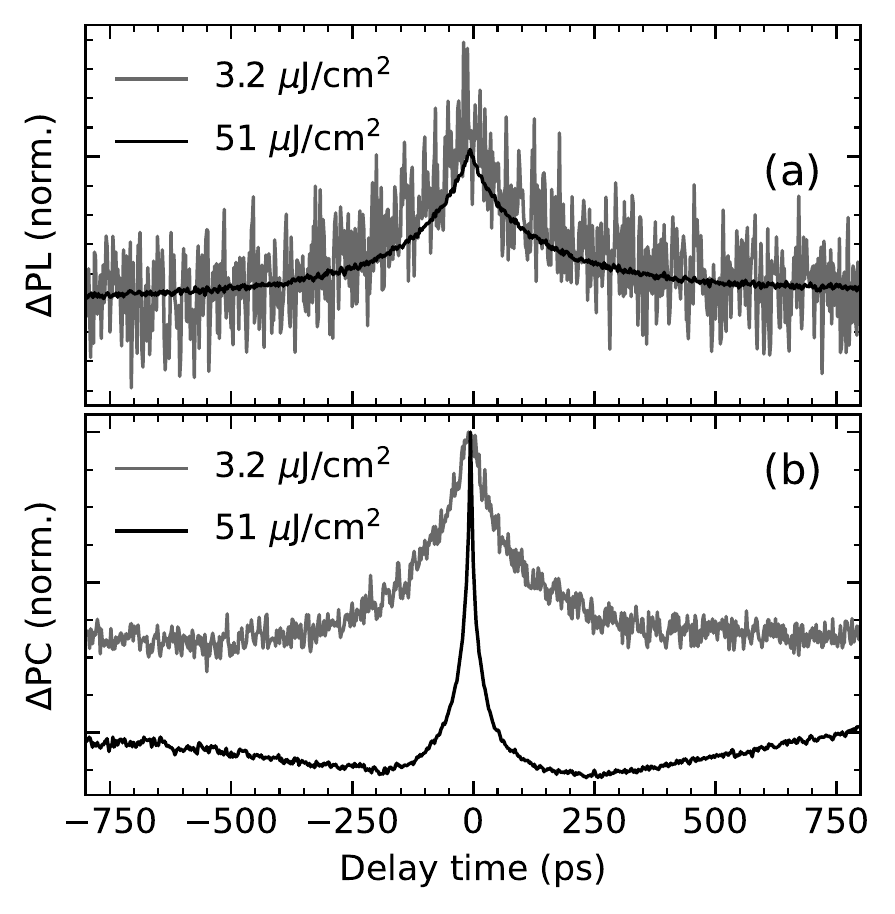}
\caption{\label{fig:epsart} Normalized nonlinear response. (a) Photoluminescence detected and (b) photocurrent detected signal.}
\end{figure}

In our simplified model, we assume that the dynamics of electron and hole carriers are comparable, which is very likely as the system is not doped.
Additionally, $\gamma_D$ is the monomolecular decay rate of the carriers and $\gamma_B$ is the non-geminate recombination rate.
Note that equation~\ref{last} is similar to equations~\ref{bimolecular} and~\ref{Anh}. From this, we can deduce that the nonlinear photocurrent (ECPC) will have a negative sign due to the non-geminate recombination experienced by the carriers and that the time trace will follow the carrier's time evolution. In this case, however, we need to consider a time-dependent generation term. We cannot assume it to be a delta function. This makes an analytical solution challenging.
\begin{equation}
    \frac{dn_{e/h}}{dt} = G(t) - \gamma_D\,n_{e/h} - \gamma_B\,n_{e/h}^2 .
    \label{last}
\end{equation}
Experimentally, in the ECPC response, shown in Figs.~\ref{ECPLOrg}(b) and~\ref{fig:epsart}(b), we observe a ``rising'' feature. We interpret this feature as the generation time of charge carriers. 
Note that as the incident fluence increases, the generation rate increases as well. Based on this experimental observation we discuss possible generation mechanisms below.

Consider the case of a charge carrier being generated through a two-step excitation, as has been proposed for several polymeric materials~\cite{Silva2001,silva2002exciton,Stevens2001,Gambetta2005}. 
For this case, since the photocarriers are directly pumped, the generation function is $G(t) \propto \delta(t)$.
We can discard this as the dominant mechanism as we expect it to manifest, in the ECPC, as a maximum in the absolute signal when the two pulses temporally overlap. Instead, experimentally we observe a minimum.
A similar experiment as the one presented here probed the two-step excitation pathway and observed that the response follows the exciton decay dynamics~\cite{Gambetta2005}. At high fluences, they observed a ``rise'' time interpreted as a fast relaxation from a hot vibrational state $S_1^*$ to $S_1$ followed by a subsequent excitation from $S_1$ to $S_n$. The rise time that we observed is too long to correspond to any relaxation time which is usually in the sub-picosecond time scale~\cite{Gambetta2005}. 

Other possible mechanisms are charge generation after bimolecular exciton-exciton annihilation~\cite{Stevens2001} and a recently proposed monomolecular exciton dissociation~\cite{Price2022}. 
In both cases, the generation rates will depend on the population of the excitons. In the bimolecular exciton-exciton annihilation pathway, the generation function would be proportional to the square of the exciton population, $G(t) \propto n(t)^2$. 
While for monomolecular exciton dissociation, the generation rate is proportional to the exciton population, $G(t)\propto n(t)$.
Notice in Fig.~\ref{ECPLOrg}(b), further exemplified in Fig.~\ref{fig:epsart}(b), that the ``rise'' of the nonlinear response is dependent on the excitation density. 
Since excitons population evolution $n(t)$ depends on the excitation density, both mechanisms show the same trend to become faster as the fluence increases. However, since we observed experimentally a more dramatic effect in the photocurrent detection scheme than in the photoluminescence scheme, shown in Fig.~\ref{fig:epsart}, we hypothesize that the dominant generation mechanism is bimolecular annihilation of the exciton population.

Furthermore, the charge lifetime observed in ECPC (Fig.~\ref{ECPLOrg}(b)) is considerably longer than the exciton lifetime (Fig.~~\ref{ECPLOrg}(a)). We cannot quantify it as it is outside of the instrument's temporal window. 
As the fluence increases, we observe that the decay of the charge becomes more significant, indicating that the bimolecular recombination of carriers becomes more important as expected from the simple model presented above. Finally, we acknowledge that there are other causes of nonlinear photocurrent that were not discussed in this work, e.g., the current limitation due to the external resistance series~\cite{Zeiske2022}. We tried to minimize these effects by performing the measurements at a low fluence range. 

\subsubsection{Summary of nonlinear dynamics in organic semiconductors}
We have shown how ECPL can be used to extract photophysical parameters like monomolecular and bimolecular decay rate constants. In the ECPC experiments we interpret the "rising" features around zero time delay as the time-dependent generation of charge carriers, which becomes faster with increasing fluence. We suggest that charge carriers are generated through bimolecular annihilation pathways. Since the fluence dependence of the generation time might originate from a second-order charge generation process, $G(t) \propto n^2$. Other mechanisms could also be involved, however, due to the strong photoluminescence nonlinearity observed, we expect bimolecular exciton-exciton annihilation to be the dominant pathway toward charge generation. To further clarify the complete photophysical scenario, systematic experiments using complementary techniques (e.g.\ transient absorption and time-resolved photoluminescence spectroscopies) are needed and recommended as future endeavors. 

\section{\label{Dis}Discussion}
In this work, we presented two scenarios in which the nonlinear PL and PC detection schemes provide distinct and complementary information.
In the case of lead-halide perovskites, the same excited-state species contribute to both the nonlinear PL and PC. 
Therefore via ECPL and ECPC, we follow the population of the same species; however, this species leads to distinct nonlinear responses in each of the physical observables.
This offers the possibility to selectively study photophysical processes experienced by the excited species based on the detection scheme. For example, we showed that while ECPL shows a response due to trap-assisted recombination, the ECPC is insensitive to traps themselves. The magnitude of ECPC, instead, is given by bimolecular recombination, a process to which ECPL is insensitive. 
As mentioned above, Zhou~\textit{et al.}~\cite{Zhou2021} have taken advantage of this to characterize carrier diffusion in layered perovskites using photocurrent detection.

For the case of organic semiconductors, we probe exclusively the excitons that recombine radiatively or those that dissociate and generate charge carriers, with each detection scheme, photoluminescence, and photocurrent respectively. In that sense,  the work presented here adds to the existing toolbox of ultrafast photoluminescence and photocurrent techniques~\cite{Bakulin2016, Bakulin2012}, with the additional feature that the magnitude of the response can be used to describe the rates responsible for the nonlinearities. In this material, both ECPL and ECPC have negative nonlinear responses which are directly related to the exciton-exciton bimolecular annihilation rate and the charge carrier bimolecular recombination respectively. As the excitonic scenarios become complex, the magnitude of the nonlinear signal provides insight into nonlinear processes occurring on the ultrafast scale hidden to steady-state measurements or that appear convoluted in time-resolved techniques.

As mentioned in the Introduction, recent reports describe a variation of the ECS probe, utilizing a tunable narrow excitation wavelength to characterize layered perovskite quantum-well structures~\cite{Zhou2019, Ouyang2020, Zhou2021}.
We note that in their interpretation there is ambiguity in the distinction between the incoherent and coherent contributions to the nonlinear response. The measured spectra are interpreted as 2D excitation spectra, but we highlight that there is no well-defined phase resolution in the excitation-pulse wavepackets, and the measurements are thus purely incoherent as in the work presented here. This incoherent response arises from the dependence of the physical observable on the intensity of the excitation due to the population evolution (e.g., trap recombination and exciton-exciton recombination), rather than a coherent nonlinear response as in coherent multidimensional spectroscopies~\cite{Tekavec2007JCP, Cundiff2013,vella2016ASciRep, Pascal2017PRB,meza2021molecular}. We also note that these 2D measurements that implement phase modulation may also contain incoherent contributions due to nonlinear population dynamics picked up by the phase demodulation detection scheme~\cite{Pascal2017, Kalaee2019,bargigia2022identifying}. 
We thus underline the difference between the technique presented in this article and 2D coherent excitation. Earlier, ECS-like experiments have been interpreted using Feynmann diagrams~\cite{Ouyang2020, Zhou2019}. 
We emphasize that this is not precise, since Feynmann diagrams indicate optical transitions among states and their coherent correlation but do not include the interactions among their populations. 
This imprecision is addressed in recent literature recognizing recombination dynamics as the only origin of the measured nonlinearity~\cite{McNamee}. 
Due to their distinct origin, they provide distinct information. 
While 2D coherent excitation experiments provide information regarding dephasing rates and coherent correlations between excited states, the ECS experiments provide information uniquely about population mixing. 
In this work, we expand on the signal-generation mechanisms associated with population mixing. Together with previous examples~\cite{VONDERLINDE1981, Rosen, Chilla1993, pau1998, Hirori2006, Miyauchi2009, Kandada2016, Perini2022, Yangwei2022}; our work adds another tool to the modern semiconductor community for the characterization of nonlinear photophysical processes.

\section{\label{per}Conclusion}

We have observed and rationalized the main nonlinear signatures in the photoluminescence and photocurrent of lead-halide perovskite and organic semiconductor devices. For the case of LHPs, the ECPL has nonlinear components due to trap-assisted and Auger recombination with opposite behavior, sublinear and supralinear. 
The fluence dependence of ECPL provides rich information about defect density, ultrafast dynamics, and Auger recombination. Meanwhile, in ECPC, the nonlinear signature originates from bimolecular and Auger recombination, both of which are supralinear processes.
In ECPC, the nonlinear contributions are convoluted and difficult to distinguish. 
In organic semiconductors, we describe ECPL as a sensitive technique for determining exciton-annihilation rates. On the other hand, ECPC represents a valuable tool to study charge generation through photo-excitation. The experimental data suggest that ECPC can follow the population dynamics of free charges including their generation dynamics. Additionally, from the rise time observed in ECPC, we hypothesize that bimolecular annihilation corresponds to a significant pathway for charge carrier generation. 

We expect EC spectroscopy to have an impact, particularly in the field of organic electronics, where it can shine further insight into the physical nature of excited states and the generation mechanisms leading to charge carriers. Besides the case of study of single component materials, mixed systems with complex fluence-dependent photophysical processes will also benefit from EC spectroscopy. For example, in recent studies of perovskite-sensitized TTA-UC, the intensity dependence of photoluminescence shows an interplay of processes with distinct nonlinearities~\cite{VanOrman2021} which could be better resolved by EC spectroscopy as well as their time-resolved dynamics. 

\section{Appendix}

\section{Excitation Correlation Spectroscopy}

In our implementation, 1030\,nm, $\sim$220 fs pulses are generated in an ultrafast laser system at a 100\,kHz repetition rate (PHAROS Model PH1-20-0200-02-10, Light Conversion). A portion of the laser beam is sent into a commercial optical parametric amplifier (ORPHEUS, Light Conversion). The pulse trains are then split 50/50 by a beam splitter cube, where one of the beams is directed to a motorized linear stage (LTS300, Thorlabs), allowing for control of the delay between the two pulses. Each pulse is modulated with a chopper at the frequencies of 373 and 199\,Hz, respectively, and the pulses are then focused onto the sample with a 100\,mm focal length lens. The total integrated response and the nonlinear component are obtained simultaneously by demodulating both the fundamental and the sum of the modulation frequency. Photoluminescence detection (ECPL): The emitted PL is filtered with a long-pass filter to get rid of the pump, and then it is focused into a photoreceiver (New Focus 2031 PR) connected to a lock-in amplifier (HF2LI, Zurich Instruments). Photocurrent detection (ECPC): The device is connected to a Zurich Instruments HF2TA Current Amplifier used to convert the current output of the sample device to voltage, as well as to supply an external bias to the device. The current amplifier is connected to a lock-in amplifier (HF2LI, Zurich Instruments). The photocurrent measurements presented here were acquired with no external applied bias. 

\subsection{Signal recovery from lock-in amplifier}

Additionally, in this appendix, we expand on the experimental details for measuring the nonlinear component utilizing double modulation lock-in detection.
The intention is to provide two examples of nonlinear photophysics processes recovered through double demodulation and to bring attention to the fiendish experimental details. 
We follow reference \citenum{Johnson1988} for the case of trap-assisted recombination. 
We define the generation rate to take into account the repetition rate and $S(t, \omega)$ to be a square wave to mimic the chopper. 
\begin{equation}
G(t,\omega) = gS(t, \omega)\sum^\infty_n\delta(t-nt_{rep})   
\end{equation}
Remember that the square wave function that alternates between 0 and 1 is given as: 
\begin{equation}
    S(t,\omega)=\frac{1}{2}+\frac{2}{\pi}\sum_{n=0}^\infty \frac{\sin((2n+1)\omega t+\theta)}{(2n+1)}.
\end{equation}

Consider the cases where the reference signal, with which the signal is demodulated, corresponds to a sine function or a square wave. Also, we will ignore the phase as this can be easily set experimentally.

\subsection{Trap-assisted recombination}
We take $\gamma=\gamma_r N_r = \gamma_t n_t$ and both pulses to have the same intensity. Then, using the equations~\ref{np} we integrate $\int^{t_{rep}}_0Bn(t)p(t)dt$ which corresponds to the response of the detector. Since $t_{rep}$ is much longer than the carrier lifetime we integrate from zero to infinity instead and obtain the intensity:

\begin{gather}
\begin{split}
    I(t, \omega_1, \omega_2) \propto \frac{S(t,\omega_1)+S(t,\omega_2)}{2}+S(t,\omega_1)S(t,\omega_2)e^{-\gamma\tau}.
\end{split}
\end{gather}

All the constants were grouped with the response of the detector.

We mimic the demodulation of the lock-in amplifier by multiplying the signal by:
\begin{equation}
    S_{ref}(t,\omega) = A \sum_{n=0}^\infty \frac{\sin((2n+1)\omega t)}{(2n+1)}.
\end{equation}. 
Then we average over a long time such that oscillating components vanish. Then the intensity recovery for each modulation frequency is:
\begin{equation}
    \langle I_{mod}(\omega_1)\rangle_{LI} \propto \frac{A}{2\pi}\sum_{n=0}^\infty\frac{1}{(2n+1)^2}\left(1+e^{-\gamma \tau}\right).
\end{equation}
\begin{equation}
    \langle I_{mod}(\omega_1)\rangle_{LI} = \langle I_{mod}(\omega_2)\rangle_{LI} \propto \frac{A\pi}{16}\left(1+e^{-\gamma \tau}\right).
\end{equation}

Note that part of the mixed term is recovered in the single modulation since $\langle S(t,\omega)\rangle = 1/2 $.
Now, we expand the mixed term to:
\begin{multline}
     S(t,\omega_1)S(t,\omega_2) = \frac{1}{4} + \frac{1}{\pi}\sum_{i=0}^2\sum_{n=0}^\infty\frac{\sin((2n+1)\omega_i t)}{(2n+1)}+\\
     \frac{4}{\pi^2}\sum_{n,m=0}^\infty\frac{\sin((2n+1)\omega_1 t)}{(2n+1)}\frac{\sin((2m+1)\omega_2 t)}{(2m+1)}.
\end{multline}

After we demodulate at the sum frequency $\omega_1+\omega_2$ and average a long time. The only terms that survive come from the last sum, when $n$ and $m$ are the same. Then:

\begin{equation}
\langle S(t,\omega_1)S(t,\omega_2)S_{ref}(t,\omega_1+\omega_2)\rangle = \frac{A}{2\pi^3}\sum^\infty_{n=1}\frac{1}{(2n+1)^3} = A\epsilon.
\end{equation}

Note that demodulating using a sine function recovers only the $n=0$ coefficient of the Fourier series.
\begin{equation}
    \langle I_{mod}(\omega_1+\omega_2)\rangle_{LI} \propto A\epsilon e^{-\gamma \tau}.
\end{equation}

\subsection{Bimolecular Annihilation}
We choose the delay between the pulses to be zero for simplicity. From the equation above, the total photoluminescence detected is:
\begin{equation}
   I_{Total\,PL} \propto \int_0^\infty n(t)dt  \propto \ln\left[1+n_i\frac{\gamma_A}{\gamma_{eff}}\right].
\end{equation}

We define $n_i = g(S(t,\omega_1)+S(t,\omega_2))$. Then, we do a second-order Taylor expansion, and $\alpha = g\gamma_A/\gamma_{eff}$. 
\begin{equation}
\approx \alpha(S(t,\omega_1)+S(t,\omega_2))-\frac{\alpha^2}{2}(S(t,\omega_1)+S(t,\omega_2))^2.
\end{equation}
Remember that the square wave is an idempotent function.
\begin{gather}
    \begin{split}
= \alpha(S(t,\omega_1)+S(t,\omega_2))\left(1-\frac{\alpha}{2}\right)-\alpha^2S(t,\omega_1)S(t,\omega_2).
    \end{split}
\end{gather}
 Then after demodulating with a square function, we obtained:
\begin{equation}
    \langle I_{mod}(\omega_1)\rangle_{LI} = \langle I_{mod}(\omega_2)\rangle_{LI} \propto \frac{A\pi}{8} \alpha\left(1-\alpha\right).
\end{equation}
\begin{equation}
    \langle I_{mod}(\omega_1+\omega_2)\rangle_{LI} \propto -A\epsilon \alpha^2.
\end{equation}

\begin{acknowledgement}
The authors thank Prof.\ Ajay Ram Srimath Kandada for fruitful discussions on the ECPL and ECPC experimental techniques and data analysis, and Victoria Quiros-Cordero for rigorous proofreading. 
The preparation of perovskite samples, devices, and their characterization and analysis, and writing of the corresponding manuscript sections by ERG, YS, DSG, and CSA was supported by the U.S. Department of Energy’s Office of Energy Efficiency and Renewable Energy (EERE) under the Solar Energy Technologies Office 24 Award Number DE-EE0008747.
The preparation of the organic-semiconductor samples was carried out by KY, supported by the Swiss National Science Foundation (200020\_184819). Their optical characterization and analysis, and writing of the corresponding manuscript sections by ERG, KY, YZ, NB, and CSA were supported by the National Science Foundation (DMR-1904293). In addition, KY acknowledges the Swiss Academy of Sciences and Universität Bern for additional travel grants. Part of this work was carried out using the shared facilities of the UW Molecular Engineering Materials Center (MEM-C), a Material Research Science and Engineering Center (DMR-1719797) supported by the U.S. National Science Foundation. 
\end{acknowledgement}

\begin{suppinfo}

See the Supplementary Information for details of the setup, solar cell, and diode used. 
Additionally, it contains details on the ECPL and ECPC models and curve-fitting parameters.

\end{suppinfo}


\begin{mcitethebibliography}{43}
\providecommand*\natexlab[1]{#1}
\providecommand*\mciteSetBstSublistMode[1]{}
\providecommand*\mciteSetBstMaxWidthForm[2]{}
\providecommand*\mciteBstWouldAddEndPuncttrue
  {\def\EndOfBibitem{\unskip.}}
\providecommand*\mciteBstWouldAddEndPunctfalse
  {\let\EndOfBibitem\relax}
\providecommand*\mciteSetBstMidEndSepPunct[3]{}
\providecommand*\mciteSetBstSublistLabelBeginEnd[3]{}
\providecommand*\EndOfBibitem{}
\mciteSetBstSublistMode{f}
\mciteSetBstMaxWidthForm{subitem}{(\alph{mcitesubitemcount})}
\mciteSetBstSublistLabelBeginEnd
  {\mcitemaxwidthsubitemform\space}
  {\relax}
  {\relax}

\bibitem[VanOrman \latin{et~al.}(2021)VanOrman, Drozdick, Wieghold, and
  Nienhaus]{VanOrman2021}
VanOrman,~Z.~A.; Drozdick,~H.~K.; Wieghold,~S.; Nienhaus,~L. Bulk halide
  perovskites as triplet sensitizers: progress and prospects in photon
  upconversion. \emph{J. Mater. Chem. C} \textbf{2021}, \emph{9},
  2685--2694\relax
\mciteBstWouldAddEndPuncttrue
\mciteSetBstMidEndSepPunct{\mcitedefaultmidpunct}
{\mcitedefaultendpunct}{\mcitedefaultseppunct}\relax
\EndOfBibitem
\bibitem[Herz \latin{et~al.}(2004)Herz, Silva, Grimsdale, M\"ullen, and
  Phillips]{Herz2004}
Herz,~L.~M.; Silva,~C.; Grimsdale,~A.~C.; M\"ullen,~K.; Phillips,~R.~T.
  Time-dependent energy transfer rates in a conjugated polymer guest-host
  system. \emph{Phys. Rev. B} \textbf{2004}, \emph{70}, 165207\relax
\mciteBstWouldAddEndPuncttrue
\mciteSetBstMidEndSepPunct{\mcitedefaultmidpunct}
{\mcitedefaultendpunct}{\mcitedefaultseppunct}\relax
\EndOfBibitem
\bibitem[Riley \latin{et~al.}(2022)Riley, Sandberg, Li, Meredith, and
  Armin]{Riley2022}
Riley,~D.~B.; Sandberg,~O.~J.; Li,~W.; Meredith,~P.; Armin,~A.
  Quasi-Steady-State Measurement of Exciton Diffusion Lengths in Organic
  Semiconductors. \emph{Phys. Rev. Appl.} \textbf{2022}, \emph{17},
  024076\relax
\mciteBstWouldAddEndPuncttrue
\mciteSetBstMidEndSepPunct{\mcitedefaultmidpunct}
{\mcitedefaultendpunct}{\mcitedefaultseppunct}\relax
\EndOfBibitem
\bibitem[Silva \latin{et~al.}(2001)Silva, Dhoot, Russell, Stevens, Arias,
  MacKenzie, Greenham, Friend, Setayesh, and M\"ullen]{Silva2001}
Silva,~C.; Dhoot,~A.~S.; Russell,~D.~M.; Stevens,~M.~A.; Arias,~A.~C.;
  MacKenzie,~J.~D.; Greenham,~N.~C.; Friend,~R.~H.; Setayesh,~S.; M\"ullen,~K.
  Efficient exciton dissociation via two-step photoexcitation in polymeric
  semiconductors. \emph{Phys. Rev. B} \textbf{2001}, \emph{64}, 125211\relax
\mciteBstWouldAddEndPuncttrue
\mciteSetBstMidEndSepPunct{\mcitedefaultmidpunct}
{\mcitedefaultendpunct}{\mcitedefaultseppunct}\relax
\EndOfBibitem
\bibitem[Daniel \latin{et~al.}(2007)Daniel, Westenhoff, Makereel, Friend,
  Beljonne, Herz, and Silva]{Clement2007}
Daniel,~C.; Westenhoff,~S.; Makereel,~F.; Friend,~R.~H.; Beljonne,~D.;
  Herz,~L.~M.; Silva,~C. Monte Carlo Simulation of Exciton Bimolecular
  Annihilation Dynamics in Supramolecular Semiconductor Architectures. \emph{J.
  Phys. Chem. C} \textbf{2007}, \emph{111}, 19111--19119\relax
\mciteBstWouldAddEndPuncttrue
\mciteSetBstMidEndSepPunct{\mcitedefaultmidpunct}
{\mcitedefaultendpunct}{\mcitedefaultseppunct}\relax
\EndOfBibitem
\bibitem[Firdaus \latin{et~al.}(2020)Firdaus, Le~Corre, Karuthedath, Liu,
  Markina, Huang, Chattopadhyay, Nahid, Nugraha, Lin, Seitkhan, Basu, Zhang,
  McCulloch, Ade, Labram, Laquai, Andrienko, Koster, and
  Anthopoulos]{Firdaus2020}
Firdaus,~Y. \latin{et~al.}  Long-range exciton diffusion in molecular
  non-fullerene acceptors. \emph{Nat. Commun.} \textbf{2020}, \emph{11},
  5220\relax
\mciteBstWouldAddEndPuncttrue
\mciteSetBstMidEndSepPunct{\mcitedefaultmidpunct}
{\mcitedefaultendpunct}{\mcitedefaultseppunct}\relax
\EndOfBibitem
\bibitem[{von der Linde} \latin{et~al.}(1981){von der Linde}, Kuhl, and
  Rosengart]{VONDERLINDE1981}
{von der Linde},~D.; Kuhl,~J.; Rosengart,~E. Picosecond correlation effects in
  the hot luminescence of \ce{GaAs}. \emph{J. Lumin} \textbf{1981},
  \emph{24-25}, 675--678\relax
\mciteBstWouldAddEndPuncttrue
\mciteSetBstMidEndSepPunct{\mcitedefaultmidpunct}
{\mcitedefaultendpunct}{\mcitedefaultseppunct}\relax
\EndOfBibitem
\bibitem[Rosen \latin{et~al.}(1981)Rosen, Doukas, Budansky, Katz, and
  Alfano]{Rosen}
Rosen,~D.; Doukas,~A.~G.; Budansky,~Y.; Katz,~A.; Alfano,~R.~R. Time resolved
  luminescence of photoexcited p‐type gallium arsenide by population mixing.
  \emph{Appl. Phys. Lett.} \textbf{1981}, \emph{39}, 935--937\relax
\mciteBstWouldAddEndPuncttrue
\mciteSetBstMidEndSepPunct{\mcitedefaultmidpunct}
{\mcitedefaultendpunct}{\mcitedefaultseppunct}\relax
\EndOfBibitem
\bibitem[Johnson \latin{et~al.}(1988)Johnson, McGill, and Hunter]{Johnson1988}
Johnson,~M.~B.; McGill,~T.~C.; Hunter,~A.~T. Picosecond time‐resolved
  photoluminescence using picosecond excitation correlation spectroscopy.
  \emph{J. Appl. Phys.} \textbf{1988}, \emph{63}, 2077--2082\relax
\mciteBstWouldAddEndPuncttrue
\mciteSetBstMidEndSepPunct{\mcitedefaultmidpunct}
{\mcitedefaultendpunct}{\mcitedefaultseppunct}\relax
\EndOfBibitem
\bibitem[Chilla \latin{et~al.}(1993)Chilla, Buccafusca, and Rocca]{Chilla1993}
Chilla,~J. L.~A.; Buccafusca,~O.; Rocca,~J.~J. Origin of photoluminescence
  signals obtained by picosecond-excitation correlation measurements.
  \emph{Phys. Rev. B} \textbf{1993}, \emph{48}, 14347--14355\relax
\mciteBstWouldAddEndPuncttrue
\mciteSetBstMidEndSepPunct{\mcitedefaultmidpunct}
{\mcitedefaultendpunct}{\mcitedefaultseppunct}\relax
\EndOfBibitem
\bibitem[Pau \latin{et~al.}(1998)Pau, Kuhl, Khan, and Sun]{pau1998}
Pau,~S.; Kuhl,~J.; Khan,~M.~A.; Sun,~C.~J. Application of
  femtosecond-excitation correlation to the study of emission dynamics in
  hexagonal \ce{GaN}. \emph{Phys. Rev. B} \textbf{1998}, \emph{58},
  12916--12919\relax
\mciteBstWouldAddEndPuncttrue
\mciteSetBstMidEndSepPunct{\mcitedefaultmidpunct}
{\mcitedefaultendpunct}{\mcitedefaultseppunct}\relax
\EndOfBibitem
\bibitem[Hirori \latin{et~al.}(2006)Hirori, Matsuda, Miyauchi, Maruyama, and
  Kanemitsu]{Hirori2006}
Hirori,~H.; Matsuda,~K.; Miyauchi,~Y.; Maruyama,~S.; Kanemitsu,~Y. Exciton
  Localization of Single-Walled Carbon Nanotubes Revealed by Femtosecond
  Excitation Correlation Spectroscopy. \emph{Phys. Rev. Lett.} \textbf{2006},
  \emph{97}, 257401\relax
\mciteBstWouldAddEndPuncttrue
\mciteSetBstMidEndSepPunct{\mcitedefaultmidpunct}
{\mcitedefaultendpunct}{\mcitedefaultseppunct}\relax
\EndOfBibitem
\bibitem[Miyauchi \latin{et~al.}(2009)Miyauchi, Matsuda, and
  Kanemitsu]{Miyauchi2009}
Miyauchi,~Y.; Matsuda,~K.; Kanemitsu,~Y. Femtosecond excitation correlation
  spectroscopy of single-walled carbon nanotubes: Analysis based on
  nonradiative multiexciton recombination processes. \emph{Phys. Rev. B}
  \textbf{2009}, \emph{80}, 235433\relax
\mciteBstWouldAddEndPuncttrue
\mciteSetBstMidEndSepPunct{\mcitedefaultmidpunct}
{\mcitedefaultendpunct}{\mcitedefaultseppunct}\relax
\EndOfBibitem
\bibitem[Srimath~Kandada \latin{et~al.}(2016)Srimath~Kandada, Neutzner,
  D’Innocenzo, Tassone, Gandini, Akkerman, Prato, Manna, Petrozza, and
  Lanzani]{Kandada2016}
Srimath~Kandada,~A.~R.; Neutzner,~S.; D’Innocenzo,~V.; Tassone,~F.;
  Gandini,~M.; Akkerman,~Q.~A.; Prato,~M.; Manna,~L.; Petrozza,~A.; Lanzani,~G.
  Nonlinear Carrier Interactions in Lead Halide Perovskites and the Role of
  Defects. \emph{J. Am. Chem. Soc.} \textbf{2016}, \emph{138},
  13604--13611\relax
\mciteBstWouldAddEndPuncttrue
\mciteSetBstMidEndSepPunct{\mcitedefaultmidpunct}
{\mcitedefaultendpunct}{\mcitedefaultseppunct}\relax
\EndOfBibitem
\bibitem[Zhou \latin{et~al.}(2019)Zhou, Hu, Ouyang, Williams, Yan, You, and
  Moran]{Zhou2019}
Zhou,~N.; Hu,~J.; Ouyang,~Z.; Williams,~O.~F.; Yan,~L.; You,~W.; Moran,~A.~M.
  Nonlinear Photocurrent Spectroscopy of Layered 2D Perovskite Quantum Wells.
  \emph{J. Phys. Chem. Lett.} \textbf{2019}, \emph{10}, 7362--7367, PMID:
  31711289\relax
\mciteBstWouldAddEndPuncttrue
\mciteSetBstMidEndSepPunct{\mcitedefaultmidpunct}
{\mcitedefaultendpunct}{\mcitedefaultseppunct}\relax
\EndOfBibitem
\bibitem[Ouyang \latin{et~al.}(2020)Ouyang, Zhou, Hu, Williams, Yan, You, and
  Moran]{Ouyang2020}
Ouyang,~Z.; Zhou,~N.; Hu,~J.; Williams,~O.~F.; Yan,~L.; You,~W.; Moran,~A.~M.
  Nonlinear fluorescence spectroscopy of layered perovskite quantum wells.
  \emph{J. Chem. Phys} \textbf{2020}, \emph{153}, 134202\relax
\mciteBstWouldAddEndPuncttrue
\mciteSetBstMidEndSepPunct{\mcitedefaultmidpunct}
{\mcitedefaultendpunct}{\mcitedefaultseppunct}\relax
\EndOfBibitem
\bibitem[Zhou \latin{et~al.}(2021)Zhou, Ouyang, Yan, McNamee, You, and
  Moran]{Zhou2021}
Zhou,~N.; Ouyang,~Z.; Yan,~L.; McNamee,~M.~G.; You,~W.; Moran,~A.~M.
  Elucidation of Quantum-Well-Specific Carrier Mobilities in Layered
  Perovskites. \emph{J. Phys. Chem. Lett.} \textbf{2021}, \emph{12},
  1116--1123\relax
\mciteBstWouldAddEndPuncttrue
\mciteSetBstMidEndSepPunct{\mcitedefaultmidpunct}
{\mcitedefaultendpunct}{\mcitedefaultseppunct}\relax
\EndOfBibitem
\bibitem[Perini \latin{et~al.}(2022)Perini, Rojas-Gatjens, Ravello,
  Castro-Mendez, Hidalgo, An, Kim, Lai, Li, Silva-Acuña, and
  Correa-Baena]{Perini2022}
Perini,~C. A.~R.; Rojas-Gatjens,~E.; Ravello,~M.; Castro-Mendez,~A.-F.;
  Hidalgo,~J.; An,~Y.; Kim,~S.; Lai,~B.; Li,~R.; Silva-Acuña,~C.;
  Correa-Baena,~J.-P. Interface Reconstruction from Ruddlesden–Popper
  Structures Impacts Stability in Lead Halide Perovskite Solar Cells.
  \emph{Advanced Materials} \textbf{2022}, \emph{34}, 2204726\relax
\mciteBstWouldAddEndPuncttrue
\mciteSetBstMidEndSepPunct{\mcitedefaultmidpunct}
{\mcitedefaultendpunct}{\mcitedefaultseppunct}\relax
\EndOfBibitem
\bibitem[Shi \latin{et~al.}(2022)Shi, Rojas-Gatjens, Wang, Pothoof,
  Giridharagopal, Ho, Jiang, Taddei, Yang, Sanehira, Irwin, Silva-Acuña, and
  Ginger]{Yangwei2022}
Shi,~Y.; Rojas-Gatjens,~E.; Wang,~J.; Pothoof,~J.; Giridharagopal,~R.; Ho,~K.;
  Jiang,~F.; Taddei,~M.; Yang,~Z.; Sanehira,~E.~M.; Irwin,~M.~D.;
  Silva-Acuña,~C.; Ginger,~D.~S. (3-Aminopropyl)trimethoxysilane Surface
  Passivation Improves Perovskite Solar Cell Performance by Reducing Surface
  Recombination Velocity. \emph{ACS Energy Letters} \textbf{2022}, \emph{7},
  4081--4088\relax
\mciteBstWouldAddEndPuncttrue
\mciteSetBstMidEndSepPunct{\mcitedefaultmidpunct}
{\mcitedefaultendpunct}{\mcitedefaultseppunct}\relax
\EndOfBibitem
\bibitem[Stranks \latin{et~al.}(2014)Stranks, Burlakov, Leijtens, Ball,
  Goriely, and Snaith]{Stranks2014}
Stranks,~S.~D.; Burlakov,~V.~M.; Leijtens,~T.; Ball,~J.~M.; Goriely,~A.;
  Snaith,~H.~J. Recombination Kinetics in Organic-Inorganic Perovskites:
  Excitons, Free Charge, and Subgap States. \emph{Phys. Rev. Applied}
  \textbf{2014}, \emph{2}, 034007\relax
\mciteBstWouldAddEndPuncttrue
\mciteSetBstMidEndSepPunct{\mcitedefaultmidpunct}
{\mcitedefaultendpunct}{\mcitedefaultseppunct}\relax
\EndOfBibitem
\bibitem[deQuilettes \latin{et~al.}(2019)deQuilettes, Frohna, Emin, Kirchartz,
  Bulovic, Ginger, and Stranks]{deQuilettes2019}
deQuilettes,~D.~W.; Frohna,~K.; Emin,~D.; Kirchartz,~T.; Bulovic,~V.;
  Ginger,~D.~S.; Stranks,~S.~D. Charge-Carrier Recombination in Halide
  Perovskites. \emph{Chem. Rev.} \textbf{2019}, \emph{119}, 11007--11019\relax
\mciteBstWouldAddEndPuncttrue
\mciteSetBstMidEndSepPunct{\mcitedefaultmidpunct}
{\mcitedefaultendpunct}{\mcitedefaultseppunct}\relax
\EndOfBibitem
\bibitem[Kiligaridis \latin{et~al.}(2021)Kiligaridis, Frantsuzov, Yangui, Seth,
  Li, An, Vaynzof, and Scheblykin]{Kiligaridis2021}
Kiligaridis,~A.; Frantsuzov,~P.~A.; Yangui,~A.; Seth,~S.; Li,~J.; An,~Q.;
  Vaynzof,~Y.; Scheblykin,~I.~G. Are Shockley-Read-Hall and ABC models valid
  for lead halide perovskites? \emph{Nat. Commun.} \textbf{2021}, \emph{12},
  3329\relax
\mciteBstWouldAddEndPuncttrue
\mciteSetBstMidEndSepPunct{\mcitedefaultmidpunct}
{\mcitedefaultendpunct}{\mcitedefaultseppunct}\relax
\EndOfBibitem
\bibitem[Ziffer \latin{et~al.}(2016)Ziffer, Mohammed, and Ginger]{Ziffer2016}
Ziffer,~M.~E.; Mohammed,~J.~C.; Ginger,~D.~S. Electroabsorption Spectroscopy
  Measurements of the Exciton Binding Energy, Electron–Hole Reduced Effective
  Mass, and Band Gap in the Perovskite CH3NH3PbI3. \emph{ACS Photonics}
  \textbf{2016}, \emph{3}, 1060--1068\relax
\mciteBstWouldAddEndPuncttrue
\mciteSetBstMidEndSepPunct{\mcitedefaultmidpunct}
{\mcitedefaultendpunct}{\mcitedefaultseppunct}\relax
\EndOfBibitem
\bibitem[Borgwardt \latin{et~al.}(2015)Borgwardt, Sippel, Eichberger, Semtsiv,
  Masselink, and Schwarzburg]{Borgwardt2015}
Borgwardt,~M.; Sippel,~P.; Eichberger,~R.; Semtsiv,~M.~P.; Masselink,~W.~T.;
  Schwarzburg,~K. Excitation correlation photoluminescence in the presence of
  Shockley-Read-Hall recombination. \emph{J. Appl. Phys.} \textbf{2015},
  \emph{117}, 215702\relax
\mciteBstWouldAddEndPuncttrue
\mciteSetBstMidEndSepPunct{\mcitedefaultmidpunct}
{\mcitedefaultendpunct}{\mcitedefaultseppunct}\relax
\EndOfBibitem
\bibitem[Paquin \latin{et~al.}(2011)Paquin, Latini, Sakowicz, Karsenti, Wang,
  Beljonne, Stingelin, and Silva]{Paquin2011}
Paquin,~F.; Latini,~G.; Sakowicz,~M.; Karsenti,~P.-L.; Wang,~L.; Beljonne,~D.;
  Stingelin,~N.; Silva,~C. Charge Separation in Semicrystalline Polymeric
  Semiconductors by Photoexcitation: Is the Mechanism Intrinsic or Extrinsic?
  \emph{Phys. Rev. Lett.} \textbf{2011}, \emph{106}, 197401\relax
\mciteBstWouldAddEndPuncttrue
\mciteSetBstMidEndSepPunct{\mcitedefaultmidpunct}
{\mcitedefaultendpunct}{\mcitedefaultseppunct}\relax
\EndOfBibitem
\bibitem[Stevens \latin{et~al.}(2001)Stevens, Silva, Russell, and
  Friend]{Stevens2001}
Stevens,~M.~A.; Silva,~C.; Russell,~D.~M.; Friend,~R.~H. Exciton dissociation
  mechanisms in the polymeric semiconductors poly(9,9-dioctylfluorene) and
  poly(9,9-dioctylfluorene-co-benzothiadiazole). \emph{Phys. Rev. B}
  \textbf{2001}, \emph{63}, 165213\relax
\mciteBstWouldAddEndPuncttrue
\mciteSetBstMidEndSepPunct{\mcitedefaultmidpunct}
{\mcitedefaultendpunct}{\mcitedefaultseppunct}\relax
\EndOfBibitem
\bibitem[Silva \latin{et~al.}(2002)Silva, Russell, Dhoot, Herz, Daniel,
  Greenham, Arias, Setayesh, M{\"u}llen, and Friend]{silva2002exciton}
Silva,~C.; Russell,~D.~M.; Dhoot,~A.~S.; Herz,~L.~M.; Daniel,~C.;
  Greenham,~N.~C.; Arias,~A.~C.; Setayesh,~S.; M{\"u}llen,~K.; Friend,~R.~H.
  Exciton and polaron dynamics in a step-ladder polymeric semiconductor: the
  influence of interchain order. \emph{Journal of Physics: Condensed Matter}
  \textbf{2002}, \emph{14}, 9803\relax
\mciteBstWouldAddEndPuncttrue
\mciteSetBstMidEndSepPunct{\mcitedefaultmidpunct}
{\mcitedefaultendpunct}{\mcitedefaultseppunct}\relax
\EndOfBibitem
\bibitem[K{\"o}hler and B{\"a}ssler(2015)K{\"o}hler, and
  B{\"a}ssler]{kohler2015electronic}
K{\"o}hler,~A.; B{\"a}ssler,~H. \emph{Electronic processes in organic
  semiconductors: An introduction}; John Wiley \& Sons, 2015\relax
\mciteBstWouldAddEndPuncttrue
\mciteSetBstMidEndSepPunct{\mcitedefaultmidpunct}
{\mcitedefaultendpunct}{\mcitedefaultseppunct}\relax
\EndOfBibitem
\bibitem[Gambetta \latin{et~al.}(2005)Gambetta, Virgili, and
  Lanzani]{Gambetta2005}
Gambetta,~A.; Virgili,~T.; Lanzani,~G. Ultrafast excitation cross-correlation
  photoconductivity in polyfluorene photodiodes. \emph{Appl. Phys. Lett.}
  \textbf{2005}, \emph{86}, 253509\relax
\mciteBstWouldAddEndPuncttrue
\mciteSetBstMidEndSepPunct{\mcitedefaultmidpunct}
{\mcitedefaultendpunct}{\mcitedefaultseppunct}\relax
\EndOfBibitem
\bibitem[Price \latin{et~al.}(2022)Price, Hume, Ilina, Wagner, Tamming, Thorn,
  Jiao, Goldingay, Conaghan, Lakhwani, Davis, Wang, Xue, Lu, Chen, Zhan, and
  Hodgkiss]{Price2022}
Price,~M.~B. \latin{et~al.}  Free charge photogeneration in a single component
  high photovoltaic efficiency organic semiconductor. \emph{Nature
  Communications} \textbf{2022}, \emph{13}, 2827\relax
\mciteBstWouldAddEndPuncttrue
\mciteSetBstMidEndSepPunct{\mcitedefaultmidpunct}
{\mcitedefaultendpunct}{\mcitedefaultseppunct}\relax
\EndOfBibitem
\bibitem[Zeiske \latin{et~al.}(2022)Zeiske, Li, Meredith, Armin, and
  Sandberg]{Zeiske2022}
Zeiske,~S.; Li,~W.; Meredith,~P.; Armin,~A.; Sandberg,~O.~J. Light intensity
  dependence of the photocurrent in organic photovoltaic devices. \emph{Cell
  Reports Physical Science} \textbf{2022}, \emph{3}, 101096\relax
\mciteBstWouldAddEndPuncttrue
\mciteSetBstMidEndSepPunct{\mcitedefaultmidpunct}
{\mcitedefaultendpunct}{\mcitedefaultseppunct}\relax
\EndOfBibitem
\bibitem[Bakulin \latin{et~al.}(2016)Bakulin, Silva, and Vella]{Bakulin2016}
Bakulin,~A.~A.; Silva,~C.; Vella,~E. Ultrafast Spectroscopy with Photocurrent
  Detection: Watching Excitonic Optoelectronic Systems at Work. \emph{J. Phys.
  Chem. Lett.} \textbf{2016}, \emph{7}, 250--258\relax
\mciteBstWouldAddEndPuncttrue
\mciteSetBstMidEndSepPunct{\mcitedefaultmidpunct}
{\mcitedefaultendpunct}{\mcitedefaultseppunct}\relax
\EndOfBibitem
\bibitem[Bakulin \latin{et~al.}(2012)Bakulin, Rao, Pavelyev, van Loosdrecht,
  Pshenichnikov, Niedzialek, Cornil, Beljonne, and Friend]{Bakulin2012}
Bakulin,~A.~A.; Rao,~A.; Pavelyev,~V.~G.; van Loosdrecht,~P. H.~M.;
  Pshenichnikov,~M.~S.; Niedzialek,~D.; Cornil,~J.; Beljonne,~D.; Friend,~R.~H.
  The Role of Driving Energy and Delocalized States for Charge Separation in
  Organic Semiconductors. \emph{Science} \textbf{2012}, \emph{335},
  1340--1344\relax
\mciteBstWouldAddEndPuncttrue
\mciteSetBstMidEndSepPunct{\mcitedefaultmidpunct}
{\mcitedefaultendpunct}{\mcitedefaultseppunct}\relax
\EndOfBibitem
\bibitem[Tekavec \latin{et~al.}(2007)Tekavec, Lott, and Marcus]{Tekavec2007JCP}
Tekavec,~P.~F.; Lott,~G.~A.; Marcus,~A.~H. Fluorescence-detected
  two-dimensional electronic coherence spectroscopy by acousto-optic phase
  modulation. \emph{J. Chem. Phys} \textbf{2007}, \emph{127}, 214307\relax
\mciteBstWouldAddEndPuncttrue
\mciteSetBstMidEndSepPunct{\mcitedefaultmidpunct}
{\mcitedefaultendpunct}{\mcitedefaultseppunct}\relax
\EndOfBibitem
\bibitem[Nardin \latin{et~al.}(2013)Nardin, Autry, Silverman, and
  Cundiff]{Cundiff2013}
Nardin,~G.; Autry,~T.~M.; Silverman,~K.~L.; Cundiff,~S.~T. Multidimensional
  coherent photocurrent spectroscopy of a semiconductor nanostructure.
  \emph{Opt. Express} \textbf{2013}, \emph{21}, 28617--28627\relax
\mciteBstWouldAddEndPuncttrue
\mciteSetBstMidEndSepPunct{\mcitedefaultmidpunct}
{\mcitedefaultendpunct}{\mcitedefaultseppunct}\relax
\EndOfBibitem
\bibitem[Vella \latin{et~al.}(2016)Vella, Li, Gr{\'e}goire, Tuladhar, Vezie,
  Few, Baz{\'a}n, Nelson, Silva-Acu{\~n}a, and Bittner]{vella2016ASciRep}
Vella,~E.; Li,~H.; Gr{\'e}goire,~P.; Tuladhar,~S.~M.; Vezie,~M.~S.; Few,~S.;
  Baz{\'a}n,~C.~M.; Nelson,~J.; Silva-Acu{\~n}a,~C.; Bittner,~E.~R. Ultrafast
  decoherence dynamics govern photocarrier generation efficiencies in polymer
  solar cells. \emph{Sci. Rep.} \textbf{2016}, \emph{6}, 1--12\relax
\mciteBstWouldAddEndPuncttrue
\mciteSetBstMidEndSepPunct{\mcitedefaultmidpunct}
{\mcitedefaultendpunct}{\mcitedefaultseppunct}\relax
\EndOfBibitem
\bibitem[Gr\'egoire \latin{et~al.}(2017)Gr\'egoire, Vella, Dyson, Baz\'an,
  Leonelli, Stingelin, Stavrinou, Bittner, and Silva]{Pascal2017PRB}
Gr\'egoire,~P.; Vella,~E.; Dyson,~M.; Baz\'an,~C.~M.; Leonelli,~R.;
  Stingelin,~N.; Stavrinou,~P.~N.; Bittner,~E.~R.; Silva,~C. Excitonic coupling
  dominates the homogeneous photoluminescence excitation linewidth in
  semicrystalline polymeric semiconductors. \emph{Phys. Rev. B} \textbf{2017},
  \emph{95}, 180201\relax
\mciteBstWouldAddEndPuncttrue
\mciteSetBstMidEndSepPunct{\mcitedefaultmidpunct}
{\mcitedefaultendpunct}{\mcitedefaultseppunct}\relax
\EndOfBibitem
\bibitem[Guti\'errez-Meza \latin{et~al.}(2021)Guti\'errez-Meza, Malatesta, Li,
  Bargigia, {Srimath Kandada}, Valverde-Ch\'avez, Kim, Li, Stingelin, Tretiak,
  Bittner, and {Silva-Acu\~na}]{meza2021molecular}
Guti\'errez-Meza,~E.; Malatesta,~R.; Li,~H.; Bargigia,~I.; {Srimath
  Kandada},~A.~R.; Valverde-Ch\'avez,~D.~A.; Kim,~S.-M.; Li,~H.; Stingelin,~N.;
  Tretiak,~S.; Bittner,~E.~R.; {Silva-Acu\~na},~C. Frenkel biexcitons in hybrid
  HJ photophysical aggregates. \emph{Sci. Adv.} \textbf{2021}, \emph{7},
  eabi5197\relax
\mciteBstWouldAddEndPuncttrue
\mciteSetBstMidEndSepPunct{\mcitedefaultmidpunct}
{\mcitedefaultendpunct}{\mcitedefaultseppunct}\relax
\EndOfBibitem
\bibitem[Grégoire \latin{et~al.}(2017)Grégoire, Srimath~Kandada, Vella, Tao,
  Leonelli, and Silva]{Pascal2017}
Grégoire,~P.; Srimath~Kandada,~A.~R.; Vella,~E.; Tao,~C.; Leonelli,~R.;
  Silva,~C. Incoherent population mixing contributions to phase-modulation
  two-dimensional coherent excitation spectra. \emph{J. Chem. Phys}
  \textbf{2017}, \emph{147}, 114201\relax
\mciteBstWouldAddEndPuncttrue
\mciteSetBstMidEndSepPunct{\mcitedefaultmidpunct}
{\mcitedefaultendpunct}{\mcitedefaultseppunct}\relax
\EndOfBibitem
\bibitem[Kalaee \latin{et~al.}(2019)Kalaee, Damtie, and Karki]{Kalaee2019}
Kalaee,~A. A.~S.; Damtie,~F.; Karki,~K.~J. {Differentiation of True Nonlinear
  and Incoherent Mixing of Linear Signals in Action-Detected 2D Spectroscopy}.
  \emph{J. Phys. Chem. A} \textbf{2019}, \emph{123}, 4119--4124\relax
\mciteBstWouldAddEndPuncttrue
\mciteSetBstMidEndSepPunct{\mcitedefaultmidpunct}
{\mcitedefaultendpunct}{\mcitedefaultseppunct}\relax
\EndOfBibitem
\bibitem[Bargigia \latin{et~al.}(2022)Bargigia, Gutiérrez-Meza,
  Valverde-Chávez, Marques, Srimath~Kandada, and
  Silva]{bargigia2022identifying}
Bargigia,~I.; Gutiérrez-Meza,~E.; Valverde-Chávez,~D.~A.; Marques,~S.~R.;
  Srimath~Kandada,~A.~R.; Silva,~C. Identifying incoherent mixing effects in
  the coherent two-dimensional photocurrent excitation spectra of
  semiconductors. \emph{J. Chem. Phys.} \textbf{2022}, \emph{157}, 204202\relax
\mciteBstWouldAddEndPuncttrue
\mciteSetBstMidEndSepPunct{\mcitedefaultmidpunct}
{\mcitedefaultendpunct}{\mcitedefaultseppunct}\relax
\EndOfBibitem
\bibitem[McNamee \latin{et~al.}(2023)McNamee, Ouyang, Yan, Gan, Zhou, Williams,
  You, and Moran]{McNamee}
McNamee,~M.~G.; Ouyang,~Z.; Yan,~L.; Gan,~Z.; Zhou,~N.; Williams,~O.~F.;
  You,~W.; Moran,~A.~M. Uncovering Transport Mechanisms in Perovskite Materials
  and Devices with Recombination-Induced Action Spectroscopies. \emph{J. Phys.
  Chem. C} \textbf{2023}, \emph{127}, 2782--2791\relax
\mciteBstWouldAddEndPuncttrue
\mciteSetBstMidEndSepPunct{\mcitedefaultmidpunct}
{\mcitedefaultendpunct}{\mcitedefaultseppunct}\relax
\EndOfBibitem
\end{mcitethebibliography}
\providecommand{\noopsort}[1]{}\providecommand{\singleletter}[1]{#1}%
\providecommand{\latin}[1]{#1}
\makeatletter
\providecommand{\doi}
  {\begingroup\let\do\@makeother\dospecials
  \catcode`\{=1 \catcode`\}=2 \doi@aux}
\providecommand{\doi@aux}[1]{\endgroup\texttt{#1}}
\makeatother
\providecommand*\mcitethebibliography{\thebibliography}
\csname @ifundefined\endcsname{endmcitethebibliography}
  {\let\endmcitethebibliography\endthebibliography}{}

\pagestyle{plain}


\section*{Supplementary material}

# Supplementary information: Resolving nonlinear dynamics in semiconductors via excitation correlation spectroscopy: Peculiar signatures of photoluminescence versus photocurrent detection

Esteban Rojas-Gatjens[1], Kaila Yallum[2], Yangwei Shi[3,4], Yulong Zheng[1], Tyler Bills[1], Carlo Andrea Riccardo Perini[5], Juan-Pablo Correa-Baena[5], David Ginger[3], Natalie Banerji[2], Carlos Silva-Acuña[‡1,5,6]

[1] *School of Chemistry and Biochemistry, Georgia Institute of Technology, 901 Atlantic Drive, Atlanta, Georgia 30332, USA.*

[2] *Department of Chemistry, Biochemistry, and Pharmaceutical Sciences, University of Bern, Freiestrasse 3, CH-3012 Bern, Switzerland*

[3] *Department of Chemistry, University of Washington, Seattle, WA 98195, USA.*

[4] *Molecular Engineering & Sciences Institute, University of Washington, Seattle, WA 98195, USA.*

[5] *School of Materials Science and Engineering, Georgia Institute of Technology, North Avenue, Atlanta, GA 30332, United States.*
E-mail: carlos.silva@gatech.edu

[6] *School of Physics, Georgia Institute of Technology, 837 State Street, Atlanta, Georgia 30332, United States.*

‡ *Currently visiting as Honorary Professor, Departamento de Física Aplicada, Centro de Investigación y de Estudios Avanzados del Instituto Politécnico Nacional, 97310 Mérida, Yucatán, México.*

## Contents

# 1 Experimental details:

### 1.0.1 ECPC discussion. Perovskite solar cell case

To ensure collection of the ECPC signal, all current transients should be completed at a speed faster than the lock-in modulation used. The response time of a solar cell under pulsed illumination is computed as:

$$t_R = \sqrt{t_{Drift}^2 + t_{Diffusion}^2 + t_{RC}^2} \tag{S1}$$

Where $t_{Drift}$ is the charge collection time for charges in the depleted region of the junction, $t_{Diffusion}$ is the collection time for charge carriers in the undepleted region, and $t_{RC}$ is the response time induced by the combination of the diode and the circuit. For a perovskite solar cell of about 1 cm$^2$ area, the intrinsic response times are significantly faster than $t_{RC}$. The response time measured can therefore be approximated as:

$$t_R = t_{RC} \tag{S2}$$

Where $t_{RC} = 2.2\,RC$, where R is the sum of the diode series and amplifier input resistances, and C is the sum of the solar cell junction and stray capacitances. In our system R $\approx$ 50 $\Omega$, the input resistance of the amplifier, and C $\approx$ 100 nF, the capacitance of the solar cell junction (assuming 1 nF/mm$^2$ area capacitance and 1 cm$^2$ area). Therefore $t_{RC} \approx 10\,\mu s$ ($10^5$ Hz), and the system response remains orders of magnitude faster than the fastest of the chopper frequencies detected by the system $\Omega_1 + \Omega_2 = 572$ Hz.

## 1.1 Sample details

### 1.1.1 Perovskite solar cell devices

For the perovskite solar cell devices, we prepared inverted devices with a mixed-cation mixed-halide perovskite of composition FA$_{0.83}$Cs$_{0.17}$Pb(I$_{0.85}$Br$_{0.15}$)$_3$ (denoted as Cs17Br15) and device architecture ITO/MeO-2PACz/Cs17Br15/C60/BCP/Ag. The Patterned ITO glass substrates were thoroughly cleaned by sonicating them in water (with 2 % Micro-90 detergent), deionized water, acetone, isopropanol(IPA) for 10 mins, respectively, followed by plasma cleaning for 5 mins. 1 mmol/L of MeO-2PACz solution was used for spin-coating on top of ITO substrates with 3000 rpm for 30 s, which was then annealed at 100 °C for 10 min. The perovskite with a concentration of 1.2 M (dissolved in DMF:DMSO = 4:1 in volume ratio) layer was spin-coated at 4000 rpm for 60 s. Chlorobenzene (CB) antisolvent was dropped on top when 35 s remained. The perovskite films were annealed at 100 degrees for 30 s and 150 °C for 10 min. After spin-coating the perovskite layer, 30 nm C$_{60}$ and 5 nm bathocuproine (BCP) were thermally evaporated, followed by 100 nm of Ag. We measured the current density-voltage (J-V) curves of the devices using a Keithley 2400 source meter under 1 Sun illumination (AM 1.5G, 100 mW/cm2) in a nitrogen glovebox. The light source was calibrated with a filtered KG3 silicon reference solar cell. The J-V curves were recorded in the range of -0.1-1.2 V with a step of 0.02 V. The solar cell devices were masked with a metal aperture (0.0453 cm2) to define the active area.

### 1.1.2 Organic single component device

The organic semiconductor devices were prepared using the non-fullerene acceptor large molecule ITIC-4F with an architecture ITO/ZnO/ITIC-4F/MoO$_3$/Ag. A 10

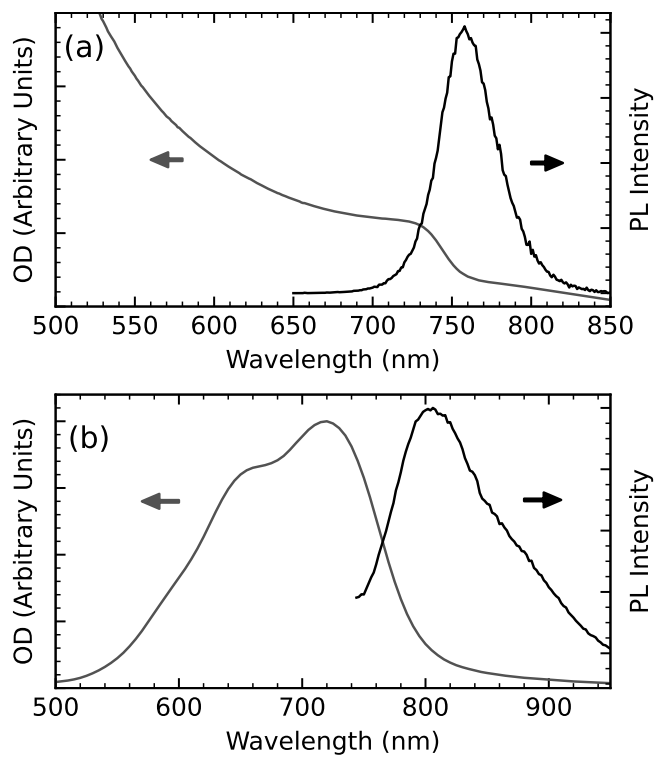

Figure S1: Absorbance and PL for (a) Cs17Br15 half-stack device (ITO/MeO-2PACz/Cs17Br15) and (b) ITIC-4F samples.

nm layer of ZnO was spincoated from 70 $\mu$L of 0.1 M ZnO nanoparticles in 2-methoxyethanol at 3000 rpm for 45 seconds. After spincoating, this layer was dried at 120° C for 10 minutes and brought to room temperature on the cooling hotplate. The active layer had a thickness of 85 nm, spincoated from 60 $\mu$L of an 8 mg/mL ITIC-4F solution in chloroform at 800 rpm for 55 seconds. MoO$_3$ was deposited by evaporation at a rate of 0.1 Å/s with an ultimate thickness of 5 nm. A 100 nm layer of Ag was deposited by evaporation at a rate of 1 Å/s.

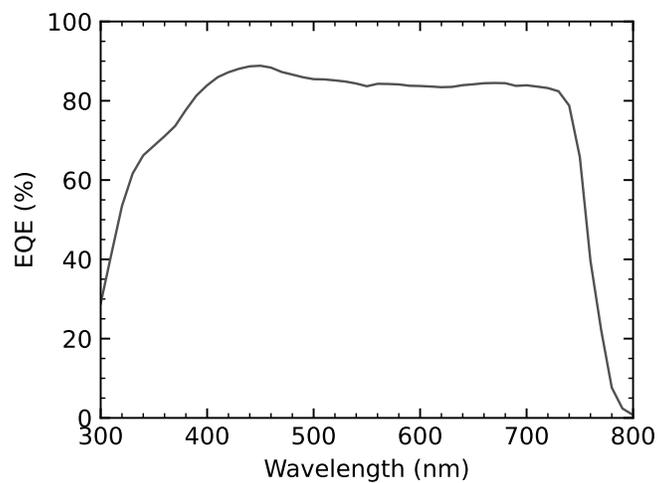

Figure S2: External quantum efficiency (EQE) of Cs17Br15 full device.

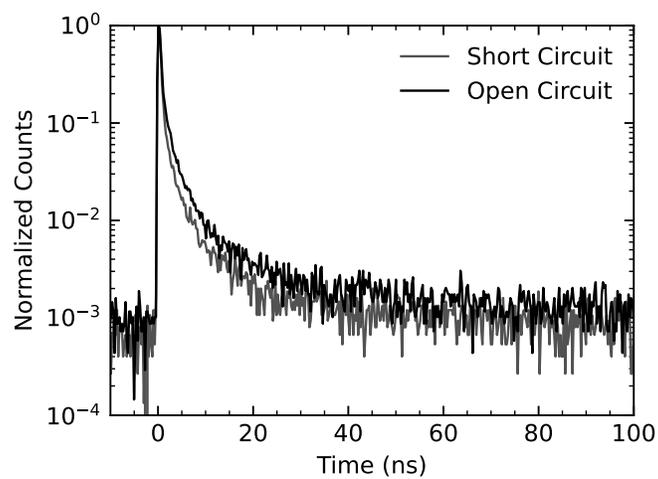

Figure S3: Time-resolved photoluminescence measurement of Cs17Br15 device in open and short circuit conditions.

## 1.2 Some simple models

### 1.2.1 Trap-assisted recombination:

ECPC:
$$n(t) = n(0)\exp(-\gamma_t N_t t) \quad \text{and} \quad p(t) = p(0)\exp(-\gamma_r n_t t). \tag{S3}$$

$$PC_{total} \propto \int_0^\tau n_1(t) + p_1(t) dt + \int_\tau^\infty n_2(t') + p_2(t') dt'. \tag{S4}$$

$$\begin{aligned} PC_{total} \propto & \int_0^\tau n(0)\exp(-\gamma_t N_t t) + p(0)\exp(-\gamma_r n_t t) dt \\ & + \int_\tau^\infty (n(\tau) + n(0))\exp(-\gamma_t N_t t') + (p(\tau) + p(0))\exp(-\gamma_r n_t t') dt'. \end{aligned} \tag{S5}$$

This is the same as

$$PC_{total} \propto 2n_0^2 + 2p_0^2 \tag{S6}$$

### 1.2.2 Bimolecular recombination:

If $Bnp >> \gamma_t N_t n$ equations 1, 2 and 3 from the main text simplify to:

$$\frac{dn}{dt} = G(t) - Bn^2 \tag{S7}$$

$$n = \frac{n_0}{1 + Bn_0 t} \quad \text{and} \quad n^2 = \frac{n_0^2}{(1 + Bn_0 t)^2} \tag{S8}$$

$$PL_{total} \propto \int_0^\tau n_1^2(t) dt + \int_\tau^\infty n_2^2(t-\tau) dt \tag{S9}$$

$$= \frac{n_0}{B}(1 - n(\tau)) + \frac{n_0}{B}(1 + n(\tau)) = 2n_0 \tag{S10}$$

The total photoluminescence corresponds to the contribution of the individual pulses. The expressions diverge in the case of photocurrent detection.

### 1.2.3 Auger recombination:

ECPL. Considering $\tau = 0$ still can be understood analytically.

$$\frac{dn}{dt} = -\gamma n - An^3. \tag{S11}$$

$$n(t) = \sqrt{\frac{\gamma/A}{(1 + \gamma/n_0^2 A)e^{2\gamma t} - 1}}. \tag{S12}$$

Integrated PL:

$$PL(n_0) \propto \int_0^\infty \frac{1}{e^2\gamma t - (1 + \gamma/n_0^2 A)^{-1}} dt \propto \left(1 + \frac{\gamma}{n_0^2 A}\right) \ln\left(\frac{\gamma + n_0^2 A}{\gamma}\right) \tag{S13}$$

The nonlinear PL
$$PL_{nl} = PL(2n_0) - 2PL(n_0) \tag{S14}$$

In the case where Auger recombination dominates $\frac{\gamma}{n_0^2 A} \to 0$ then:

$$PL_{nl} = \ln\left(\frac{\gamma + 4n_0^2 A}{\gamma}\right) - 2\ln\left(\frac{\gamma + n_0^2 A}{\gamma}\right) \tag{S15}$$

$$= \ln\left(\frac{\gamma^2 + 4n_0^2 A\gamma}{\gamma^2 + 2n_0^2 A\gamma + n_0^4 A^2}\right) \tag{S16}$$

The previous expression is negative if $\frac{\gamma}{n_0^2 A} < 1/2$, which is true based on our previous assumptions.

ECPC. We simplify to a steady state case due to the complex showing the sign of the signal. $\gamma n \ll An^3$. Then the nonlinear signal is clearly negative.

$$\frac{dn}{dt} = 0 = G - An^3 \tag{S17}$$

$$PC_{nl} = \frac{\left(2^{1/3} - 2\right) G^{1/3}}{A} \tag{S18}$$

## 1.3 Fitting procedures

### 1.3.1 Quasi-Steady

Solving equation 14 one obtains:

$$n(t) = \frac{n_0 \gamma/\beta}{(n_0 + \gamma/\beta)\exp(\gamma t) - n_0} \tag{S19}$$

Since we are excited with a single pulse then the total photoluminescence measure is just the integral of the expression above, where R groups the photodiode response and the sample radiative response.

$$I_{PL} = R \int_0^\infty \frac{n_0 \gamma/\beta}{(n_0 + \gamma/\beta)\exp(\gamma t) - n_0} = \frac{1}{\beta}\ln(1 + n_0\beta/\gamma) \tag{S20}$$

By measuring a fluence dependence of the $I_{PL}/n_0$ we can extract the ratio $\beta/\gamma$ and from the previously determined $\gamma$ we isolated the $\beta$.

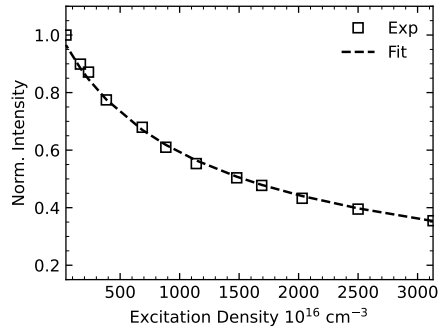

Figure S4: Caption

### 1.3.2 ECPL

The fits using Equation 15 are displayed in Figure S5, where the fit parameters required are shown in Table 1. The measurement at fluence 1.0 $\mu J/cm^2$ is shown here for demonstration purposes but the fit parameters are ignored due to the signal being dominated by noise.

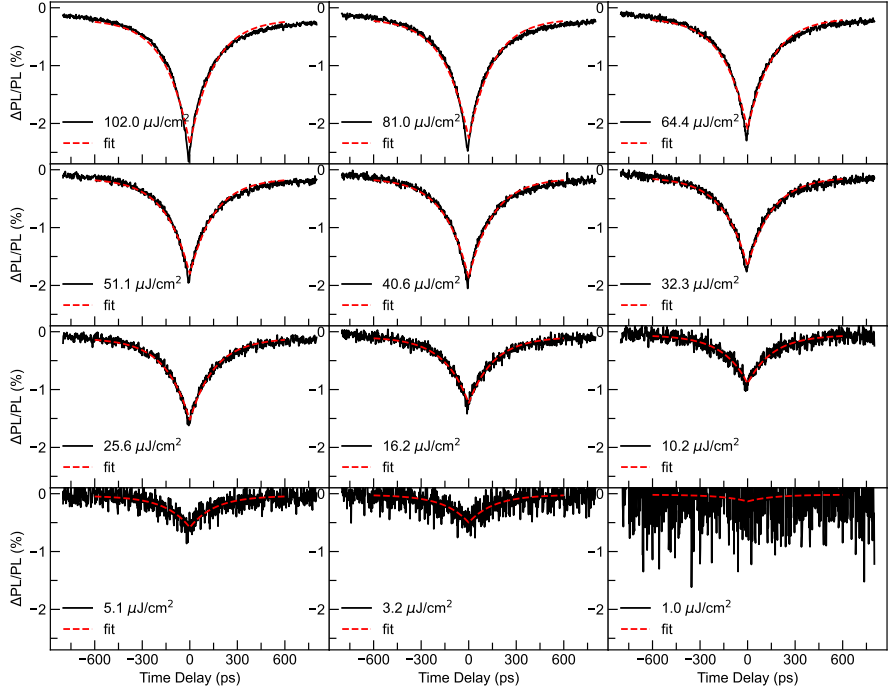

Figure S5: Fits (red dashed line) using equation 14 in the main article.

### 1.3.3 ECPC

We fit only the right arm of the respective time traces. For the three lowest fluences we consider either a single or a double rising exponential, which is sufficient to describe the dynamics. We assign no physical value to the double exponential other than to report an average rise time. As the fluence increases a decay in the signal can be observed and then we incorporate a single exponential. A general expression is shown in equation S21.

$$f(t) = A(1 - B\exp(-t/\tau_{r1}) - C\exp(-t/\tau_{r2})) + D\exp(-t/\tau_d) \qquad (S21)$$

We also note that in figure 3.b, at the highest fluences the decay is very small and therefore the estimation of the lifetime is not reliable. Instead we focus on analysing the rise time constants. The results of the fits are summarized in table 2

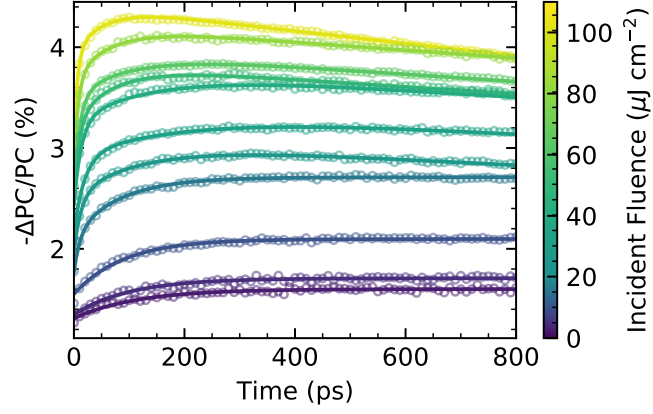

Figure S6: Fitting to a single rising exponential of the ECPC response of ITIC-4F for the three lowest fluences.

Table 1: Summary of the extracted photophysical parameters for ITIC-4F from the ECPL presented in figure ??.a of the main text.

| Fluence ($\mu$J cm$^{-2}$) | $\alpha$ | $\gamma_{bim}$ ($10^{-23}$ cm$^3$ ps$^{-1}$) | offset (%) |
|---|---|---|---|
| 3 | $0.0097 \pm 0.0005$ | $22 \pm 3$ | 0 |
| 5 | $0.0111 \pm 0.0004$ | $16 \pm 1$ | 0 |
| 10 | $0.0171 \pm 0.0003$ | $12 \pm 1$ | -0.1 |
| 16 | $0.0235 \pm 0.0002$ | $11 \pm 0.2$ | -0.1 |
| 25 | $0.0292 \pm 0.0002$ | $8.4 \pm 0.1$ | -0.1 |
| 32 | $0.0318 \pm 0.0002$ | $7.3 \pm 0.1$ | -0.1 |
| 40 | $0.0357 \pm 0.0002$ | $6.6 \pm 0.1$ | -0.1 |
| 51 | $0.0343 \pm 0.0002$ | $5.0 \pm 0.1$ | -0.1 |
| 64 | $0.0401 \pm 0.0002$ | $4.6 \pm 0.1$ | -0.2 |
| 81 | $0.0434 \pm 0.0002$ | $4.0 \pm 0.1$ | -0.2 |
| 102 | $0.0453 \pm 0.0003$ | $3.3 \pm 0.0$ | -0.2 |

Table 2: Summary of the extracted photophysical parameters for ITIC-4F from the ECPC measurements (figure 2.b of the main text).

| Fluence ($\mu$J cm$^{-2}$) | $\tau_{r1}$ (ps) | $\tau_{r2}$ (ps) | $\tau_{avg}$ (ps) |
|---|---|---|---|
| 3 | 113 $\pm$ 4 | - | 113 $\pm$ 4 |
| 5 | 102 $\pm$ 4 | - | 102 $\pm$ 4 |
| 10 | 91 $\pm$ 2 | - | 91 $\pm$ 2 |
| 16 | 84 $\pm$ 2 | 11 $\pm$ 2 | 56 $\pm$ 2 |
| 25 | 120 $\pm$ 4 | 13 $\pm$ 2 | 66 $\pm$ 4 |
| 32 | 196 $\pm$ 10 | 18 $\pm$ 4 | 89 $\pm$ 10 |
| 40 | 138 $\pm$ 5 | 13 $\pm$ 1 | 62 $\pm$ 5 |
| 51 | 70 $\pm$ 4 | 9 $\pm$ 2 | 37 $\pm$ 4 |
| 64 | 97 $\pm$ 5 | 11 $\pm$ 1 | 40 $\pm$ 5 |
| 81 | 66 $\pm$ 5 | 8 $\pm$ 2 | 30 $\pm$ 5 |
| 102 | 11 $\pm$ 5 | 6 $\pm$ 2 | 22 $\pm$ 5 |

# Graphical TOC Entry

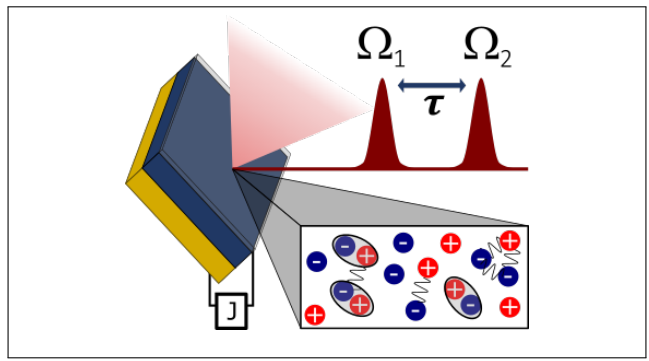



\end{document}